\documentstyle[multicol,pre,aps,epsfig,amsmath]{revtex}

\newcommand{\colrule}{\hline}

\newcommand{\dt}{\partial_t}
\newcommand{\dx}{\partial_x}
\newcommand{\dy}{\partial_y}
\newcommand{\dz}{\partial_z}
\newcommand{\druck}[0]{P\relax}
\newcommand{\fred}{Fre{\'e}dericksz\relax}
\newcommand{\nen}{b}
\newcommand{\adfiNull}{{\Phi^+_u}}
\newcommand{\adfiEins}{{\Phi^+_c}}
\newcommand{\adfiZwei}{{\Phi^+_s}}
\newcommand{\adnz}{{n_z^+}}
\newcommand{\EE}[1]{\times 10^{#1}}
\newcommand{\N}{N}
\newcommand{\pN}{}

\begin{document}

\draft

\title{Three-dimensional pattern formation, multiple homogeneous
  soft modes, and nonlinear dielectric electroconvection}
\author{Axel G. Rossberg\\
Department of Physics, Kyoto University, Kyoto 606-8502, Japan\\
http://www.rossberg.net/ag}
\date{submitted to PRE}
\maketitle
% insert suggested PACS numbers in braces on next line
%
\begin{abstract}
  Patterns forming spontaneously in extended, three-dimensional,
  dissipative systems are likely to excite several homogeneous soft
  modes ($\approx$ hydrodynamic modes) of the underlying physical
  system, much more than quasi one- and two-dimensional patterns are.
  The reason is the lack of damping boundaries.  This paper compares
  two analytic techniques to derive the patten dynamics from
  hydrodynamics, which are usually equivalent but lead to different
  results when applied to multiple homogeneous soft modes.  Dielectric
  electroconvection in nematic liquid crystals is introduced as a
  model for three-dimensional pattern formation.  The 3D pattern
  dynamics including soft modes are derived.  For slabs of large but
  finite thickness the description is reduced further to a
  two-dimensional one.  It is argued that the range of validity of 2D
  descriptions is limited to a very small region above threshold.  The
  transition from 2D to 3D pattern dynamics is discussed.
  Experimentally testable predictions for the stable range of ideal
  patterns and the electric Nusselt numbers are made.  For most
  results analytic approximations in terms of material parameters are
  given.
\end{abstract}

\begin{multicols}{2} 
\narrowtext

\section{Introduction}
\label{sec:introduction}

Spontaneous formation of spatially periodic
structures on a homogeneous background is ubiquious in nature,
fascinating to look at, and often hard to understand in detail.  The
periodic structures are almost never ideal.  Irregularities may be
generated in the transient after the pattern formation is initiated
and anneal after some time.  Or, in particular in dissipative systems
far from equilibrium, they may be the result of an instability of the
regular, spatially periodic state itself, then often leading to a
state which exhibits persistent spatio-temporally chaotic dynamics.
In some systems this is the case arbitrary close to the threshold of
pattern formation in control parameter space.
With the help of reduced descriptions like phase-diffusion and
Ginzburg-Landau like amplitude equations, which are to some extent universal
(i.e., independent of physical details), several phenomena associated
with these deviations from the simple periodic structure can be
explained \cite{croho}.

For mostly practical reasons, experimental and theoretical research on
pattern formation and dynamics has concentrated on quasi one- or
two-dimensional systems, but most of the results obtained should have
a direct correspondence also in genuinely three dimensional patterns.
By a genuinely three dimensional pattern (below simply 3D pattern) I
do here mean a spatially periodic structure for which (i) the spatial
period(s) are \textit{not} determined by the spatial extension of the
sample (referred to as the class of patterns formed by ``competing
interaction'' in Ref.~\cite{croho}) \textit{and} (ii) for which the
spatial extension of the sample is in all directions large compared to
the period(s) of the pattern.  This is a stronger conception of a ``3D
pattern'' than the one used in Refs.~\cite{juknotob,jukno}, which is
based only on (ii).

Additional complications in 3D patterns, as compared to 1D or 2D
patterns, arise from the structure and dynamics of defects
(dislocations as well as disclinations), which are point like in 2D
but line like in 3D.  The implications for dissipative, nonpotential
systems, mainly those described by the complex Ginzburg-Landau
equation, have been addressed by several authors
\cite{keener,friri,gaottgu,tornschro,arbikra,roucharay}. %,glghpl,plpe
But there is another particularity of 3D patterns, which has so far
found little attention: the massive occurrence of \emph{homogeneous
  soft modes}, which couple to the pattern and can drastically change
its dynamics.

By homogeneous soft modes I mean marginally stable or slowly decaying
homogeneous or long-wavelength perturbations of the homogeneous
basic state from which the pattern arises.  In the abstract sense of
the word, they are hydrodynamic modes of the basic state.  But for the
sake of clarity the terms ``hydrodynamic mode'' and ``hydrodynamics''
shall here be reserved for slowly relaxing deviations from the
\emph{thermodynamic} equilibrium in an unbounded, homogeneous medium,
and their dynamics (e.g., the velocity field in a convective flow is a
hydrodynamic variable).  The pattern-forming basic state is itself a
non-equilibriums state.  Thus, although there is some correspondence
between homogeneous soft modes and hydrodynamic modes (in the narrow
sense), the notions are not identical.

As it has become clear by the investigation of several 1D and 2D model
systems, homogeneous soft modes are the key for understanding many of
the phenomena occurring at, or close to, the onset of pattern formation.
The most prominent example is the mode associated with a homogeneous
perturbation of the pressure field in Rayleigh-B{\'e}nard convection
which leads to a ``singular mean flow''.  It is worth noticing that,
since it is usually possible to construct self-consistent amplitude
equations which do \emph{not} include the effect of homogeneous soft
modes, their relevance is easily underestimated in the theoretical
analysis.
  
In 1D and 2D pattern-forming systems, most hydrodynamic modes are
damped by the boundaries enclosing the system.  For example, momentum
and heat can usually diffuse freely through the boundaries and are
stabilized by large external reservoirs.  Obviously, this mechanism is
ineffective in systems which are extended in all three spatial
dimensions.  On the other hand, some coupling to a reservoir will also
be required in 3D in order to sustain non-equilibrium pattern
formation.  This could be through electromagnetic fields, some matrix
embedding the active, pattern-forming medium, or some chemical
reactant provided in excess.  But the couplings to the reservoirs are
highly specific in these cases and stabilize only a few hydrodynamic
variables.  The remaining fields do then lead to homogeneous soft
modes.  \emph{As a result, several homogeneous soft modes should be
considered as the rule in 3D, pattern-forming systems.}

For example, when studying the 3D structures formed by chemical waves
in the Belousov-Zhabotinsky (BZ) reaction, the dynamics of the plain BZ
reagent does also involve convective fluid motion.  Since these
hydrodynamic modes are usually considered to be a nuisance, they are
suppressed by embedding the reagent in a gel \cite{jahnke88:_chemic}.
But another homogeneous soft mode excited by the pattern, the
temperature field (gradients of which are probably driving the
convection) remains.  Since temperature gradients have a strong
influence on the dynamics of the pattern \cite{vinson97:_contr}, a
complete description of the 3D BZ reaction should explicitly involve
this mode.

The work presented here is a case study of 3D pattern formation in the
dielectric regime of electroconvection (EC) in nematic liquid
crystals.  The system was chosen because of its easy experimental
accessibility.  In particular, the electric nature of the instability
allows to obtain patterns with several hundred periods extension in
cells of a fingernail's size, evolving on the time scale of seconds.
A closely related variant, the conduction regime of EC, which always
leads to quasi 2D patterns, is currently one of the best understood
experimental pattern-forming systems, on the phenomenological as well
as on the quantitative level (see the reviews \cite{pebe,ehcbook}).
These advantages compensate the inconvenience of dealing with rather
complicated (electro-)hydrodynamic equations.

Rather than trying to understand the complex pattern dynamics itself,
this paper is mainly devoted to the development of consistent reduced
descriptions of the dynamics.  Section~\ref{sec:dielectric} sketches
the experimental phenomenon and the hydrodynamics of dielectric EC,
emphasizing its 3D nature.  Approximations used for an analytic or
semi-analytic description of dielectric EC are introduced in
Section~\ref{sec:linear}, thereby discussing the linear stability
problem.  In Section~\ref{sec:ddd} the 3D amplitude formalism for
dielectric EC is derived.  Close to the threshold of EC and in a
liquid-crystal slab of large but finite thickness, the pattern
dynamics becomes essentially 2D.  The corresponding equations of
motion are derived from the 3D formalism in Section~\ref{sec:dd}.  In
Section~\ref{sec:stability} the stability of ideal periodic patterns
is investigated and in Section~\ref{sec:transitions} a general
scenario for the transition from the onset of dielectric EC to fully
3D pattern dynamics with increasing external stress is developed.
Section~\ref{sec:nusselt_numbers} discusses possible experiments based
on electric Nusselt number measurements and
Section~\ref{sec:conclusion} summarizes the results.
Appendix~\ref{sec:coefficients} contains some analytic and numerical
results for coupling coefficients, Appendix~\ref{sec:methods} compares
two different methods for integrating multiple, homogeneous soft modes
into the amplitude formalism in a general framework; one method is
used in the main text.

\section{Some phenomenology of electroconvection}
\label{sec:dielectric}

Notice that below some points are oversimplified in order to ease
intuition.  For comprehensive reviews of EC see
Refs.~\cite{pebe,ehcbook,annrev}, for introductions into
nemato-hydrodynamics Refs.~\cite{dg,chl}.

\subsection{Basic phenomena}
\label{sec:basicec}

In the typical experiment a nematic liquid crystal with negative
dielectric anisotropy is sandwiched between a pair of transparent,
parallel electrodes (separation $d\sim20-50 \mu \rm m$, area $\sim
1{\rm cm}^2$).  By a special treatment of the electrode surfaces, the
nematic director $\vec n$ (the locally averaged molecular orientation;
$|\vec n|=1$) is forced to align parallel to the electrodes in some
preferred direction which shall here be identified with the
$x$-direction ($z$ be normal to the electrodes, $y$ normal to $x$ and
$z$).  An ac voltage $E_0 \, \hat z \, d \,\cos \omega t$ is applied
at the electrodes.  In the \emph{conduction regime} at frequencies
below the \emph{cut-off frequency} $\omega_c$, the first instability
to be observed as the voltage is increased is towards a pattern of
convection rolls called Williams domains \cite{willi}.

At higher frequencies a different kind of structure periodic along $x$
is found.  Compared to Williams domains it has shorter wavelength and
decays faster after switching of the voltage (fast turnoff mode).  At
least two concurring mechanisms have been proposed for this high
frequency mode: the \emph{dielectric EC} \cite{dvgpp}, which depends
essentially on the anisotropy of the nematic (its threshold diverges
at the nematic-isotropic phase transition \cite{ridu}), and the
\emph{isotropic mechanism} \cite{chipi,bablpitr} where the liquid
crystal's anisotropy is not essential for the convection mechanism
itself but only for selecting a preferred modulation direction.  It
has a finite threshold at the nematic-isotropic phase transition as
its characteristic signature \cite{bbgt}.  The two linear modes have
the same symmetry and do in principle mix, but generally the
corresponding thresholds can be assumed to be sufficiently separated
to consider the mechanisms isolatedly.  The isotropic mode is thought
to be located mainly near the electrodes, while the dielectric mode is
maximal at mid plane.  Unfortunately, it is not always clear which
mode is actually observed.  At least in some cases the dielectric mode
could be identified by the good match of the threshold curve with
theoretical predictions (e.g.\ \cite{scheukp}).  The isotropic
mechanisms will not be considered here.

For voltages slightly higher than the threshold of dielectric EC, the
formation of the \emph{chevron} superstructure is observed: defects
(dislocations) in the pattern of convection rolls accumulate along
lines oriented in $y$ direction, such that the topological charge of
the defects alternates from line to line.  Between the lines, the
convection rolls are rotated and the nematic director is twisted,
alternately clock- and counterclockwise \cite{asr}.  The observation
of chevron patterns in the conduction regime of EC with homeotropic
director alignment \cite{tobupeka,huhika2} shows that this scenario is
not restricted to a particular convection mechanism.

\subsection{Hydrodynamic equations and material parameters}
\label{sec:material}

EC in both the conduction and the dielectric regime result from the
interaction of electric field, space charges, mass flow, and the
nematic director \textit{via} the Carr-Helfrich \cite{he} mechanism:
Spatial modulations of the director orientation are amplified by an
inhomogeneous mass flow generated by electric volume forces on space
charges which accumulate due to inhomogeneous electric currents in the
inhomogeneous director field.  Thus Maxwell's equations \footnote{In
  MKSA units.}  (in the quasi-static approximation \footnote{We will
  only allow for homogeneous magnetic fields.}  $\textrm{curl} \vec
E=\textrm{curl} \vec H=0$) and the balance equations for charge,
momentum, mass (continuity equation), and the torque acting on $\vec
n$ have to be taken into account.  They contain several material
parameters: the conductivities $\sigma_\parallel \gtrsim
\sigma_\perp=O(10^{-9} \cdots 10^{-5}\, \Omega^{-1}\,\textrm{m}^{-1})$
for electric currents parallel ($\parallel$) and perpendicular
($\perp$) to $\vec n$ respectively (they vary on a large range
depending on purity and doping, while $\sigma_\parallel/\sigma_\perp$
changes only little), the dielectric constants \footnote{For
  convenience, the factor $\epsilon_0$ is absorbed into
  $\epsilon_\parallel$ and $\epsilon_\perp$.}
$\epsilon_\parallel,\epsilon_\perp=O(\epsilon_0)$ (the quantities
$\sigma_a:=\sigma_\parallel-\sigma_\perp = O(\sigma_\perp)$,
$\epsilon_a:=\epsilon_\parallel-\epsilon_\perp = O(\epsilon_\perp)$
measure their anisotropies), the flexoelectric constants
$e_1,e_3=O(10^{-12}-10^{-11}\, \textrm{C}\,\textrm{m}^{-1})$
($e_+:=e_1+e_2$, $e_-:=e_1-e_2$), the diffusion constants for (ionic)
charge carriers $O(10^{-11}\, \textrm{m}^2\,\textrm{s}^{-1})=:D_\rho$,
the mass density
$\rho_{\textrm{m}}=O(10^3\,\textrm{kg}\,\textrm{m}^{-3})$, the five
independent viscosities
$\alpha_1,...,\alpha_5=O(0.1\,\textrm{N}\,\textrm{m}^{-2}\,\textrm{s})$
($\alpha_6=\alpha_2+\alpha_3+\alpha_5$, $\gamma_1=\alpha_3-\alpha_2$,
$\gamma_2=\alpha_3+\alpha_2$, $2 \eta_1=-\alpha_2+\alpha_4+\alpha_5$,
$\eta_2=\gamma_2+\eta_1$), and the curvature elasticities of the
director field $k_{22} \lesssim k_{11} \lesssim
k_{33}=O(10^{-11}\,\textrm{N})$.  I also include the ``dynamic
flexoelectric effect'', which was predicted \cite{pleibrabook,brpl84}
on the basis of a systematic rederivation of nematohydrodynamics, but,
to the authors knowledge, has not been detected, yet.  It is
characterized by a parameter $\zeta^E$ and leads to additional
dissipative contributions in the charge, momentum, and torque balance
equations.

\subsection{Dimensional analysis}
\label{sec:dimensional}

With the exception of the charge relaxation time $\tau_0:=
\epsilon_\perp/\sigma_\perp=O(10^{-6}-10^{-1} \, \textrm{s})$ the
nematohydrodynamic equations (without external fields), being derived
as a limit of large time and length scales (though typically valid
down to molecular scales), do not set any time or length scale by
themselves.  Instead, one finds basically three types of
diffusivities: for charge ($D_\rho$), director orientation [e.g.
$D_{d,\text{stat}}=k_{33}/\gamma_1=
O(10^{-10}\,\textrm{m}^2\,\textrm{s}^{-1})$ for static,
$D_{d,\text{dyn}}=(k_{33} \eta_1)/(\gamma_1\eta_1-\alpha_2^2) =
O(10^{-9}\,\textrm{m}^2\,\textrm{s}^{-1})$ for dynamic deformations;
notice that (static) flexoelectric effects do not introduce a new
diffusive scale since $e_{1/3}^2/\epsilon_0\lesssim k_{33}$], and
momentum [e.g.\ $D_p=(\gamma_1\eta_1-\alpha_2^2)/(\gamma_1
\rho_m)=O(10^{-5}\,\textrm{m}^2\,\textrm{s}^{-1})$ along $\vec n$].
The overdamped limit $\rho_m\to0$, $D_p\to \infty$ is generally a good
approximation. Charge diffusion is not essential for the Carr-Helfrich
mechanism and is typically screened out.  Then orientational diffusion
sets the only diffusive scale.
    
With an externally generated electric ac field $E_0 \hat z \cos \omega
t$ two additional time scales are introduced: The period $2 \pi
\omega^{-1}$ and the ``director time''
$\tau_d=\gamma_1/(\epsilon_\perp E_0^2)$ [the more intuitive choice
$\tau_d:=\gamma_1/(\epsilon_a E_0^2)$ would suggest that
$\epsilon_a=0$ is singular for EC, which is, for the convective modes
themselves, not the case].
        
In the conduction regime the charge densities oscillate with the
frequency of the applied field, while the director orientation is
mostly constant.  This imposes a condition 
\begin{align}
  \label{cond-range}
  \tau_d \gtrsim \omega^{-1}
\gtrsim \tau_0
\end{align}
on the three time scales.  The finite sample thickness $d$ determines
the wavelength $\lambda$ of the convection pattern and leads through a
condition $d \sim \lambda \sim (D_{d,\text{stat}}\,\tau_d)^{1/2}$ to a
voltage threshold $V_c^2=d^2 E_c^2 \sim k_{33}/\epsilon_\perp$ for the
onset of convection (see Ref.~\cite{ehcbook} for a good analytic
formula).  A lower limit for the sample thickness is given
\textit{via} relation~(\ref{cond-range}) by $d^2 \gtrsim
D_{d,\text{stat}} \tau_0$.  At frequencies higher then the cutoff
frequency $\omega_c \sim \tau_0^{-1}$ the conduction mechanism is also
disabled or at least supersede by the dielectric mode.

In the dielectric regime director and fluid flow oscillate with
$\omega$, which leads to a condition
\begin{equation}
  \label{diel-range}
  \tau_d \lesssim \omega^{-1}.
\end{equation}
The charge distribution is at high enough frequencies
($\omega^{-1}\lesssim\tau_0$) mostly constant in time. This is not
actually necessary for the dielectric mechanism to be effective
\cite{smigaladvdu}, but typically dielectric EC is supersede by the
conductive mode at lower $\omega$.  The threshold for the onset of EC
is now given by a condition $\tau_d \sim \omega^{-1}$ (or $E_c^2
\epsilon_\perp/\gamma_1\sim\omega$), i.e., the lowest $E_0$ compatible
with relation~(\ref{diel-range}).  The wavelength $\lambda$ of the
critical mode can under some conditions be $\sim d$
\cite{smigaladvdu}, but for typical materials used it is $\lambda \sim
(D_{d,\text{dyn}} /\omega)^{1/2}$, at least as long as this length is
smaller than $d$ and larger than the Debye screening length
$(D_\rho\,\tau_0)^{1/2}$, i.e., $\omega
\tau_0<D_{d,\text{dyn}}/D_\rho=O(10^{2})$, where charge diffusion
becomes important.

Thus, the length scales given by the spatial period of the pattern
$\lambda$ and the sample thickness $d$ are usually independent and
easily separated ($\lambda \ll d$) in the dielectric regime, either by
increasing $d$ or by simultaneously increasing $\omega$ and the
conductivities, while leaving the secondary control
parameter $\omega\tau_0$ constant.  Since strong doping may affect the
nematic material parameters and the nematodynamics at high frequencies
is not fully understood, the program carried out below is best
seen as the \emph{limit of thick cells}.  The theory should accurately
describe typical experiments in cells with $d\gtrsim 10\,\lambda$.  To
observe fully three dimensional patterns, thicker cells might be
required (see also Section~\ref{sec:transitions}).

\section{Approximation methods and linear theory}
\label{sec:linear}

\subsection{2D vs. 3D amplitude formalism}
\label{sec:2Dvs3D}

Below we will develop the amplitude formalism for the pattern dynamics
in the dielectric regime, i.e., obtain the laws of motion of amplitude
and phase of the spatial modulations as described by the complex pattern
amplitudes $A^\prime(x,y,t)$ or $A(x,y,z,t)$, respectively.

The basic state in the experimental cell is anisotropic and inversion
symmetric and the primary bifurcation is supercritical (forward)
towards a steady-state pattern with a single critical wave vector.
Hence, the most elementary description of the pattern dynamics is give
by the time-dependent Ginzburg-Landau equation in 2D,
\begin{equation}
  \label{rgle}
  \tau \dt A^\prime=\left(\epsilon^\prime+ \xi_x^2 \dx^2 +\xi_y^2
    \dy^2-g^\prime |A^\prime|^2\right) A^\prime,
\end{equation}
important physical properties of which are reviewed in
Refs.~\cite{dakr,mathbook}.  The real, positive coefficients $\tau$,
$\xi_{x}$, $\xi_{y}$, and $g^\prime$ have magnitudes corresponding to
natural scales of the system (e.g.\ the pattern wavelength for
$\xi_{x}$, $\xi_{y}$) and can be calculated from the underlying
hydrodynamic equations. The small, dimensionless parameter
$\epsilon^\prime$ measures the distance from the threshold of pattern
formation in the control-parameter space of the underlying system.
    
It will be shown in Sec.~\ref{sec:dd} that, as a direct consequence of
the separation of length scales in the dielectric regime, the range of
validity of Eq.~(\ref{rgle}) is highly restricted.  Already for values
of $\epsilon^\prime$ of the order $\lambda^4/d^4$ corrections to
Eq.~(\ref{rgle}) must be taken into account.  For
$\epsilon^\prime\sim\lambda^2/d^2$ the 2D description breaks down
completely.  But the convective dynamics can then still be described
in terms of the 3D modulations of the complex pattern amplitude
$A(x,y,z,t)$ (defined by Eq.~(\ref{normalize}) below), which is
coupled nonlinearly to several homogeneous soft modes.  It is
therefore natural to derive first the 3D amplitude dynamics which can
then be reduced further to a 2D description in a subsequent step.

Julien, Knobloch, and Tobias \cite{juknotob,jukno} were the first to
implement the idea of deriving a reduced description for the
$z$-dependence of the amplitude of patterns with $\lambda/d \ll 1$ as
an intermediate step in the theory, and also the first to observe that
this method significantly eases the restriction of the control
parameter to values close to threshold.  Their calculation do,
however, not involve in-plane modulations of the pattern and the
resulting excitation of homogeneous soft modes.  Several results
concerning the 3D description of dielectric EC and its reduction to 2D
are derived in an unpublished work by Lindner \cite{linddpl}, which is
quoted here whenever necessary.

\subsection{Linear stability in 3D}
\label{sec:linear3d}

The starting point for setting up the 3D amplitude equations is to
calculate the linear threshold $E_c$, critical wave number $q_c$ and
critical eigenvector (i.e., {}the $2\pi/\omega$-periodic time
dependence of the hydrodynamic fields at threshold) ignoring any
spatial variations along $z$.  Several linear stability calculations
of this type have been carried out
\cite{dvgpp,smigaladvdu,mara,ehcbook}.
    
In experiments, the critical wave vector is always found to be
parallel to the orientation of the nematic director in the basic state
(the $x$ direction). The linear problem is thus effectively
one-dimensional, with trivial, sinusoidal variations along the
remaining $x$ direction, and is much easier to solve than the 2D
problem including variations and boundary conditions along $z$.  This
technical advantage of the 3D approach, which is of course not
restricted to EC, remains effective also in the subsequent
calculations of the coupling coefficients in the 3D amplitude equation.
    
For analytic as well as numerical calculations it is convenient to use
a truncated Fourier expansion of the time dependences of the
hydrodynamic fields, assuming them to be $2\pi/\omega$ periodic.  In
the simplest cases truncated at lowest order (i.e., including constant
and $\sin$/$\cos \omega t$ contributions) \cite{ehcbook,linddpl} or
including the $\sin$/$\cos 2 \omega t$ modulation of the induced
electric potential in order to better model the interplay between
electric charges and fields.  With these truncations, the stability
problem for sinusoidal excitation can be solved explicitly (see
Appendix~\ref{sec:coefficients}).  Notice, however, that this ``lowest
order'' or, respectively, ``second lowest order Fourier
approximation'' involves some arbitrariness in the choice of
variables and does not correspond to any physical limit.  Numerical
convergence ($5\%$ accuracy) requires inclusion of at least the third
harmonic.  The actual time dependence of, e.g., the director field,
depends on $\omega$ and is non-trivial even as $\omega \to \infty$
\cite{dvgpp}.

Some results presented here rely on the first or second lowest order
Fourier expansion of the induced electric potential $\Phi$, the
nematic director expressed by $n_z$ and $\varphi$ such that $\vec
n=n_z \hat z+(1-n_z^2)^{1/2} \hat c$, $\hat c:=(\cos \varphi,\sin
\varphi,0)$, the velocity field $\vec v$, and the pressure $\druck$.
The other hydrodynamic fields are treated implicitly.  For the
representation of the dielectric mode itself, $\vec v$ is expressed in
the divergence-free form $\vec v=(-\dy,\dx,0)\,
g+(\dx\dz,\dy\dz,-\dx^2-\dy^2)\,f$ and the pressure is eliminated.

As a natural consequence of $\tau_d \sim \omega^{-1}$, the relative
phases of electric, director and velocity fields in the linear
eigenvector are shifted by angles $O(1)$.  Remarkably, the phase shift
between the lowest Fourier mode of director oscillations and external
field is $\pi/4$, for the first \cite{linddpl} and second-lowest-order
Fourier approximation (see Appendix~\ref{sec:coefficients}) exactly and
only slightly perturbed ($<1\%$ for $\omega\tau_0>2$) when higher
Fourier modes are included.  No simple physical explanation for this
result should be expected, since it holds only at the critical (most
unstable) wavenumber.  Experimental observations \cite{stzr} seem to
agree with a value $\pi/4$ for the phase shift even better than the
comparison with Galerkin calculations including $z$-dependence
\cite{bzk}, which had been carried out as a test of the dielectric
model of EC.

We define the pattern amplitude $A$ such that the amplitude $n_{z,c}$
of the $\cos \omega t$ Fourier component of the director tilt
oscillations (which is in phase with the applied voltage) has a
spatial dependence
\begin{align}
  \label{normalize}
  n_{z,c}=A(x,y,z,t) \exp(i q_c x)+c.c\,.
\end{align}
    
Assuming as usual $\epsilon$, $\dx$, $\dy$, $\dz$ to be small and
discarding contributions beyond the lowest nontrivial order, the
linear part of the 3D amplitude equation assumes the form
\begin{equation}
  \label{3da}
  \tau \dt A=\left(\epsilon+ \xi_x^2 \dx^2 +\xi_y^2
    \dy^2 + \xi_z^2 \dz^2 \right) A.
\end{equation}
As conventional, $\epsilon:=(E_0^2-E_c^2)/E_c^2$.  The coherence
lengths $\xi_{x}$, $\xi_{y}$, $\xi_{z}$ turn out to be~$\sim \lambda$
and the relaxation time $\tau$ is of the order of the charge
relaxation time $\tau_0$  \cite{smigaladvdu} (s.
Appendix~\ref{sec:coefficients}).
      
The horizontal boundary conditions for $A$ are simply
\begin{equation}
  \label{zboundary}
  A=0 \quad \text{at} \quad z=\pm d/2.
\end{equation}
Contributions from derivatives of $A$ and nonlinear contributions to
the boundary conditions are of higher order and can be discarded.  In
particular, as is well known, the distinction between
free and no-slip boundary conditions for the velocity field plays no
role at this point.  With the realistic no-slip boundary conditions
for the velocities, the relative magnitude of the hydrodynamic fields
in the 2D linear eigenvector (including $x$ and $z$ variations)
locally deviates from the 1D eigenvector (only $x$ variation) only in
a boundary layer of thickness $\sim \lambda$, an example of which is
shown in Fig.~\ref{fig:randschicht}.  This boundary layer might
provide a problem for numerical approaches directly using 2D
eigenvectors in thick cells with no-slip boundaries, notably when
Galerkin approximations are used.

\subsection{Linear stability in a cell of finite thickness}
\label{sec:linear2d}

Assuming as usual the lateral ($x$, $y$) extensions of the cell to be
large compared to its thickness $d$, the trivial solution $A \equiv 0$
of Eq.~(\ref{3da}) with boundary conditions~(\ref{zboundary}) becomes
unstable at $\epsilon=\epsilon_d:=(\pi \xi_{z}/d)^2$ [i.e.,
$\epsilon^\prime=\epsilon-\epsilon_d$ in Eq.~(\ref{rgle})] with a
critical mode $A \sim \cos(\pi z/d)$.  The small threshold shift $E_c
\to (1+\epsilon_d/2) E_c$ due to the $z$-variation is rather
uninteresting by itself (there is also a shift $\sim \lambda^2/d^2$ in
$q_c$ by a discarded contribution to Eq.~(\ref{3da}) of the form $i
\dx \dz^2 A$), but the effect provides, for example, a simple
interpretation of the small gap ($\Delta E_0=3 \epsilon_d E_c/2$)
between the critical mode and the lowest $z$-antisymmetric mode
(i.e.~$A\sim\sin(2 \pi z/d)$) reported in \cite{kaz}.

\subsection{Flexoelectric effects}
\label{sec:flexo}

A short remark about flexoelectric effects, which are generally
difficult to isolate experimentally, is in place at this point.  The
high symmetry of the linear problem in 3D does not allow flexoelectric
effects: $E_c$, $q_c$, $\tau$, and $\xi_{x}$ are independent of the
flexoelectric coefficients.  Most of the remaining linear and
nonlinear coefficients contain flexoelectric contributions.  With our
choice of variables these contributions are, except for some
dynamic-flexoelectric terms, all indirect: flexoelectric effects
excite additional, ``slaved'' contributions in the subspace orthogonal
to the critical eigenvector, which, again by flexoelectric effects,
feed back into the dynamics of the amplitude of the eigenvector.  In
contrast, the contributions not depending on flexoelectric
coefficients are, except for $\xi_x$, all direct: no excitation of
slaved degrees of freedom is involved.  Therefore \emph{flexoelectric
  effects are separated in a natural way} from the standard dynamics.
This might provide methods for measuring the flexoelectric
coefficients in a way not sensitive to parasitic boundary effects.
Previous calculations involving flexoelectric effects
\cite{mara,tzk,kbpt} where restricted to the conventional, ``static''
flexoelectric contributions and concentrated on the determination of
critical mode and voltage.

\section{The 3D amplitude equations}
\label{sec:ddd}

\subsection{Method}
\label{sec:softmodes}

Before discussing the homogeneous soft modes relevant for dielectric
EC, some comments on methodology are required.

Multiple homogeneous soft modes excited by a patterning mode have, to
my knowledge, first been introduced by Plaut and Pesch \cite{plpe}.
But their description requires the soft-mode amplitudes to be constant
along all but one spatial direction.  This limitation seems to be
partly due to the procedure by which the equations were derived.
There are two popular philosophies for this procedure (see below and
Appendix~\ref{sec:methods}), which shall here be labeled as ``order
parameter'' method and ``center manifold'' method.   
% The two methods lead usually (for at most one
% homogeneous soft mode) to the same result, and hence their
% distinctness finds little attention.
The two methods usually (for at most one homogeneous soft mode) differ
only in the way in which the problem is formulated and solved, but
lead to the same results.  The association of existing general
prescriptions for deriving amplitude equations with the former
(e.g.~\cite{ha,cro80,nplrev,fdpk,mathbook}) and the latter philosophy
(e.g.~\cite{kured,chgooo}, the Chapman-Enskog approach
\cite{chapcobook,hydrored} for the derivation of hydrodynamics from
statistical mechanics is also of this type, see Refs.
\cite{kamp,kuramoto90}) is therefore not always conclusive (see
also~\cite{dakr,manbook,iphyso}).  When using the ``order parameter''
method to obtain reduced equations, each spatial Fourier mode of the
physical state is projected onto the (adjoint) slowly decaying linear
eigenmodes of the basic state \emph{with the corresponding wave
  vector}.  Using the ``center manifold'' method, the projection is
always onto the (adjoint) eigenvectors for (typically neutrally
stable) \emph{homogeneous perturbations}.  When there are multiple
slow modes at a single wave vector (usually $\vec q=0$), the resulting
reduced equations differ, as is shown in Appendix~\ref{sec:methods}.
Plaut and Pesch~\cite{plpe} seem to be using the ``order parameter''
method, which leads to problems in more than one spatial dimension.
Here, the ``center manifold'' method is used to derive the nonlinear
extensions of the amplitude equation~(\ref{3da}) including homogeneous
soft modes.

\subsection{Derivation of the soft mode equations}
\label{sec:smderive}

Some particularities of the problem under consideration have to be
taken into account: The quasi-static approximation of electrodynamics
$\mathop{\rm curl} \vec E=0=\mathop{\rm curl} \vec B$ and the
approximation of an incompressible fluid $\nabla \vec v=0$ both lead
to additional homogeneous ``soft modes'': Since no time derivatives of
electric potential (electric field) or pressure occur in the basic
equations as they are used here, all their temporal Fourier modes are
in the kernel of the linear operator $L(0,0)$ (s.
Appendix~\ref{sec:methods}).  When the viscid limit $\rho_m\to0$
becomes effective, the same applies for the oscillating part of the
velocity field.  This is the case when spatial variations occur on
scales smaller than $(\omega
D_p)^{1/2}=(D_p/D_{\text{dyn}})^{1/2}\,\lambda =O(10^{5/2}) \lambda$.
Formally we shall assume length scales to be larger than this.  But,
since, at least in this case, velocity and pressure oscillations do
not feed back into the remaining dynamics at lowest order in the
derivatives, i.e., there are no contributions $O(\rho_m^{-1})$, the
description should be good also on smaller scales.

To simplify the problem further, the equations for the soft modes are
here calculated only in the lowest order Fourier approximation.  The
``slaved'' modes being eliminated are then the slowly varying average
director tilt $n_z$, which is stabilized by the applied electric field
through the dielectric anisotropy $\epsilon_a$ (assumed to be negative
hereafter), and oscillations of the director, which are viscously
damped.  Finally, in anticipation of corresponding boundary
conditions, only small deviations from the basic state $\vec
v,\varphi,n_z,\Phi=0$ shall be considered for now.

In order to obtain a consistent truncation of the soft-mode equations,
recall that the only non-diffusive scale in the hydrodynamic equations
is the charge relaxation time $\tau_0$.  Assume $\omega\tau_0$ to be
fixed.  This also determines $E_c\sim\omega^{1/2}$ for given $\tau_0$
and we will assume $E_0 \approx E_c$.  The elimination the fast modes
($n_z$ and oscillations of $\vec n$) becomes more efficient when
$\omega$ and $E_0$ increase or, respectively, $\tau_0$ decreases.
Then, in the limit of small $\tau_0$ (or large $\sigma_{\parallel}$,
with fixed $\sigma_\parallel/\sigma_\perp$), the time scale $\tau_0$
drops out of the equations.  A purely diffusive scaling for the
derivatives ($\dx^2\sim\dy^2\sim\dz^2\sim\dt$) is retained, without
making any \textit{a priori} assumptions about the actual scaling laws
of typical lengths and times, which may be different.  This
approximation breaks down when length scales become shorter than
$(D_{d,{\mathrm{stat}}}\tau_0)^{1/2}$ or
$(D_{d,{\mathrm{stat}}}/\omega)^{1/2}$.

It turns out that with these approximations the only \emph{relevant} modes in
the fast subspace (i.e.\ {}$R$ in Appendix~\ref{sec:cmmethod}) are
$n_z$ and its temporal modulations as given by
\begin{equation}
  \begin{split}
    \label{nz}
    n_z&(x,y,z,t)=
    \\& 2\dz\frac{
      \left( {k_{{22}}} - {k_{{11}}} \right) {{\partial}_y}{n_y} + 
      {{\alpha}_3}{v_x}}{{E_0^2}{{\epsilon}_a}}
    \\+&
    2 \dx \frac{ {{\epsilon}_a}{E_0}{{\Phi}_r} - 
      \left({e_+}-2{{\gamma}_1}{{\zeta}^E}\right){{\partial}_z}{{\Phi}_0} 
      + {{\alpha}_2}{v_z}
      }
    {{E_0^2}{{\epsilon}_a}}
%     \\&+-
%     {\frac{2\left( {{\partial}_y}{{\partial}_z}
%           \left( -{k_{{11}}} + {k_{{22}}} \right) {n_y} + 
%           {{\alpha}_3}{{\partial}_z}{v_x} + 
%           {{\partial}_x}\left( {E_0}{{\epsilon}_a}{{\Phi}_r} - 
%             {{\partial}_z}{{\Phi}_0}{e_+} + {{\alpha}_2}{v_z} + 
%             2{{\partial}_z}{{\Phi}_0}{{\gamma}_1}{{\zeta}^E} \right) 
%         \right) }{{{{E_0}}^2}{{\epsilon}_a}}}
    \\+&  
    {\frac{2 \nen 
        \left( {{\epsilon}_a}{{\partial}_x}{{\Phi}_0} - 
          {e_-}{{\partial}_y}{n_y} \right) 
        }{E_0^3 \epsilon_a^2}}\times
    \\&\qquad\qquad \left( {E_0^2}{{\epsilon}_a}\cos (\omega t) - 
      4{{\gamma}_1}\omega\sin (\omega t) \right)
    .
  \end{split}
\end{equation}
The electric potential has been decomposed as $\Phi=\Phi_0+2 \Phi_r
\cos \omega t-2 \Phi_i \sin \omega t$.  The parameter $
\nen:={{{E_0^4}\,{{\epsilon}_a^2}}/( {3\,{E_0^4}\,{{\epsilon}_a^2} +
    16\,{{\gamma}_1^2}\,{\omega^2}}}) $ measures the strength of the
excitation of the oscillatory part of $n_z$.  It is numerically small
[in the standard material MBBA
($p$-methoxybenzilidene-$p^{\prime}$-$n$-butylaniline)
$\nen\approx0.01$]. The resulting description for the soft mode
dynamics is given by the Eq.~(\ref{mfeqs}) on page \pageref{mfeqs1}
(for the terms containing $A$, see Section~\ref{sec:interaction}
below).

\subsection{Comments on the soft mode equations}
\label{sec:smdiscuss}

Most of the terms in Eq.~(\ref{mfeqs}) reproduce linearized
nematohydrodynamics.  Equations~(\ref{fi0}-\ref{fii}) derive from the
charge balance equation [Eqs.~(\ref{fir},\ref{fii}) have been
multiplied with $E_0$], Eq.~(\ref{ny}) from the angular momentum
balance on $\vec n$, Eqs.~(\ref{vx}-\ref{vz}) from the Navier-Stokes
equation and Eq.~(\ref{p}) is the unchanged continuity equation.  In
Eq.~(\ref{ny}) a term $\chi_a H_y^2 \varphi$ has been included, which
describes the action of a magnetic field in $y$ direction.  It will be
used in Section~\ref{sec:dd}.  The contributions resulting from the
elimination of fast modes are underlined.

Remarkably, the equations are mostly independent of the strength of
the external field (a factor $E_0$ can be absorbed into the
definitions of $\Phi_r$ and $\Phi_i$), although it is the cause for
the excitations of the slaved modes.  As an example, consider the
mechanism for the reduction of viscosity by the term $\alpha_2 \dx^2
v_z$ in Eq.~(\ref{vz}) (recall $\alpha_2<0$): Shear forces
$\sim\alpha_2\dx v_z$ excite $n_z$.  This leads to polarization
charges $\sim \dx E_0 \epsilon_a n_z$.  The electric field $\sim E_0$
acting on these charges generates bulk forces on the fluid.  On the
other hand, the excitation of $n_z$ is damped by electric forces $\sim
\epsilon_a E_0^2$ and the factor $E_0^2$ cancels out.

The soft mode equations reflect the non-equilibrium character of the
basic state.  For example, if Onsager's relations would hold, the
coefficients of $\dx\dz v_x$ in Eq.~(\ref{vz}) and of $\dx\dz v_z$ in
Eq.~(\ref{vx}) would be the same.  At low $\omega$ the basic state is
even unstable. The mechanism corresponds to EC in the conduction
regime.  Ignoring flexoelectric effects, the cutoff frequency
$\omega_c$ above which the basic state stabilizes is given by
\begin{equation}
  \label{omegac}
  \omega_c^2=
  {\frac{{{\sigma}_{\parallel}}\,
      \left( {{\alpha}_2}\,{{\epsilon}_{\perp}}\,
        {{\sigma}_{\parallel}} -
        {{\alpha}_2}\,{{\epsilon}_{{\parallel}}}\,{{\sigma}_{\perp}} - 
        {{\epsilon}_a}\,{{\eta}_1}\,{{\sigma}_{\perp}} \right) }{
      {{\epsilon}_a}\,{{\epsilon}_{{\parallel}}}\,{{\epsilon}_{\perp}}\,
      {{\eta}_1}}},
\end{equation}
which reproduces the result of direct stability calculations using
(effectively) the same Fourier truncation \cite{dvgpp}.  Only the
threshold field for the Williams domains is too small to be resolved
by Eq.~(\ref{mfeqs}).

\subsection{Nonlinear extensions}
\label{sec:nonlinear_extensions}

Of course, constant values can always be added to any of the soft
modes by a Galilei transformation, a rotation, or a gauge
transformation.  The problem of adding nonlinear contributions (e.g.\ 
{}advection terms) to Eqs.~(\ref{mfeqs}) such that they become
formally invariant under these transformation is easily solved.  The
solution is not unique, but it can be seen by inspection that the
precise form of the nonlinearities does not matter under the following
conditions:
\begin{enumerate}
\item Dynamics is such that, in fact, diffusive scaling holds.  In
  particular this implies that, if $L \gg \lambda$ is the typical
  length scale (i.e. $\dx,\dy,\dz\sim L^{-1}$), the variations of
  $\vec v$, $\Phi_r$, $\Phi_i$, $\varphi/L$, and $\Phi_0/L$ over $L$
  scale in the same manner as $L\to\infty$.
\item \label{small_phi} Variations in $\varphi$ over $L$ are much
  smaller than one.  It is not necessary to specify the scaling
  relation of $\varphi$ and $L$.  When $L$ is determined through the
  bulk dynamics, $\varphi$ may actually vary by $O(1)$ over the
  sample.
\end{enumerate}
It should be noticed that, although the second condition is satisfied
for many problems of pattern dynamics, it is too strong for
disclinations (line defects) in the director field at any distance $R$
from the core of the disclination: On the typical length scale $L=R$
variations of $\varphi$ are $O(1)$.  Then, for example, nontrivial
nonlinear contributions from the velocity field $\vec v\sim L^{-1}$,
like a term of the form $(\dx v_z)(\dy \varphi)$ in Eq.~(\ref{vz}),
might have to be included.

For the part describing the curvature elasticity in Eq.~(\ref{ny}),
the fully nonlinear corrections in $\varphi$ have been calculated by
extending the ``center manifold'' method to nonlinear contributions
and requiring rotation invariance.  The result has the
same form as close to equilibrium:
\begin{multline}
  \label{nlcurvature}
  \gamma_1 \dt \varphi=k_{33} \hat c_\perp \cdot (\dx^2+\dy^2)
  \hat c \\
  +
  (k_{11}^\prime-k_{33}) (\hat c_\perp \cdot \nabla)(\nabla \cdot \hat c)+
  k_{22} \hat c_\perp \cdot \dz^2 \hat c + \ldots
\end{multline}
[$\hat c_\perp:=(-\sin \varphi,\cos \varphi,0)$], however, with
$k_{11}^\prime:={k_{{11}}}+2\, \nen\, e_-^2/\epsilon_a$.  Although an
additional term proportional to $(\nabla \cdot \hat c_\perp)(\nabla
\cdot \hat c)$ would be thinkable, it does not occur in the present
approximation.  For a rotationally invariant description of the
dynamics of the convection pattern, it is useful to go over to a
representation in terms of $\tilde A(\vec r)=A(\vec r) \exp(i q_c x)$
as in Ref.~\cite{RoKr}.

\subsection{Stability of the twist mode with respect to $x$ modulations
}
\label{sec:no_prewavye}

Another point to notice is that below the threshold of dielectric EC
(i.e.~with $A=0$) the system can, in our approximation, never
destabilize in such a way that $\varphi$ is excited but $\dy
\varphi=0$, even when allowing for flexoelectric effects and arbitrary
$z$ dependencies.  Without $y$ modulations, the in-plane director
couples only to $v_x$, and this interaction is not affected by the
elimination of the fast modes and hence relaxational.  This is
remarkable because such modulation instabilities of $\varphi$ below
the EC threshold are apparently \textit{observed in experiments}
\cite{ridu,brubliba} (the ``inertial mode'' \cite{pichi} which was
proposed as an explanation has the wrong symmetry).  A far-fetched but
possible explanation would be that the modulations in $\varphi$ are
generated through a mechanism which is similar to the one that
generates the chevron superstructure in the dielectric regime,
however, invoked by the convection rolls of ``isotropic'' EC (see
Section~\ref{sec:basicec}).  The isotropic mechanism is expected to
become active in the respective experimental situations but does
itself not involve excitations of $\varphi$.  The convection rolls
might themselves be smaller than the optical resolution and therefore
remain unobserved.  Then the onset of $\varphi$ modulation would
\emph{appear} to be the threshold of a primary instability of the
homogeneous basic state, although the actual primary (``isotropic'')
threshold is at slightly lower voltages.

\subsection{Interaction with the pattern}
\label{sec:interaction}

The equation of motion for the pattern amplitude itself is given by
\begin{multline}
  \label{full3da}
  \tau ({{{\partial}_t}} {+ \vec v \cdot\! \nabla\!+i v_x q_c}){A}=
  \left[\kappa_x \nabla_x v_x + \kappa_z \nabla_z v_z
    \vphantom{- \frac{\nabla_z {\Phi}_r}{{E_{{0}}}} } 
  \right. 
  \\
  +\xi_x^2 {\partial}_x^2 
  +\xi_y^2 ({\partial}_y^2-2 i q_c \varphi \dy- q_c^2 \varphi^2)
  +\xi_z^2 {\partial}_z^2 
  \\
    + i \alpha_x \nabla_x \Phi_0 
    + i \beta_y \nabla_y  \varphi
  \left. 
    {- \frac{4 \nabla_z
        {\Phi}_r}{{E_{{c}}}}} 
    {+ \varepsilon} 
    {- g |A|^2}
  \right] A .
\end{multline}
All coefficients are real.  The differential operators $\nabla_x$,
$\nabla_y$, $\nabla_z$ are used to indicate that only the immediately
following expression should be differentiated, not $A$.  With the
exceptions of $v_x\dx A$ and $\nabla_x v_x A$, which are included for
symmetry, only terms up to lowest nontrivial order have been included,
assuming time, all lengths and all fields to scale independently.  The
l.h.s.\ is given by Galilean invariance, the form of the expression
following $\xi_y^2$ from rotation invariance \cite{rohekrpe}.  The
term involving $\nabla_z\,\Phi_r$ represents simply an additional
contribution to the electric driving field.  Symmetry would also allow
a term $\kappa_y \nabla_y \, v_y A$, but the hydrodynamic equations do
not generate it.

When $\beta_y=-\xi_y^2 q_c$ it is possible to absorb the corresponding
term into the parenthesis after $\xi_y^2$ and to rewrite the complete
expression in the ``potential'' form $\xi_y^2
(\dy-iq_c\varphi)(\dy-iq_c\varphi) A$.  In this case the phase of the
pattern does not drift in a weakly deformed $\varphi$ field.  In fact,
$\beta_y$ is close to this value (see \cite{linddpl},
Appendix~\ref{sec:coefficients}), which is largely due to the
conservation of charge and momentum and the flux-divergence form of
the resulting expressions.

The Landau coefficient $g$ is of order unity in our normalization
\cite{linddpl}.  As in the conduction regime \cite{plpe}, it is
dominated by ``geometric'' effects, i.e., inhibition of the
Carr-Helfrich mechanism for large $n_z$: besides flexoelectric
effects, about 99\% of $g$ come from contributions quadratic or cubic
in $n_z$.  This indicates that a breakdown of the weakly nonlinear
expansion should be expected for $A\sim n_z=O(1)$.

The nonlinear excitation of the soft modes by the convection pattern
is less intuitive than their feedback onto the rolls discussed above.
The principle of truncation for the contributions of $A$ in
Eq.~(\ref{mfeqs}) is again to keep only terms of lowest nontrivial
order, however, allowing for phase gradients of $A$ without gradients
of the modulus.  As the relative scaling of $\varphi$ \textit{vs.}\ 
{}length scales is left undetermined at this stage (see
Section~\ref{sec:nonlinear_extensions}), the nonlinear term $-A^* i
q_c \varphi A$, which is given by rotation symmetry, must always come
along with $A^*\dy A$.

Since Eq.~(\ref{fi0}) must have flux-divergence form, the largest
contributions from $A$ are $O(\nabla^2 A^2)$, which is too small to be
relevant.  The coefficients $I_r$ and $I_i$ in
Eqs.~(\ref{fir},\ref{fii}) have the dimensions of a current density
and measure the strength of an alternating current $\vec j_{|A|^2}=2
|A|^2 (I_r \cos \omega t- I_i \sin \omega t)$ generated by the
convection pattern.  Important contributions to the coefficients
$I_r$,\ldots,$I_{iz}$ come from charge advection.

The strongest contributions to $\Gamma$ in Eq.~(\ref{ny}) are simple
potential effects.  In MBBA, the most important one describes a
relaxation of the bend of the director modulations by twist and is
given by $4(k_{22}-k_{33})$ (see \cite{linddpl,plpe} and
Appendix~\ref{sec:coefficients}), which is generally negative.  The
second most important contribution comes from the dielectric torques
from applied and induced field on the director.  In MBBA this
contribution to $\Gamma$ is positive and in materials with large
negative $\epsilon_a$ (e.g.\ $\lesssim -\epsilon_\perp$) it could
compensate the elastic one and reverse the sign of $\Gamma$.  Notice
also that $\Gamma$ is paricularly sensitive to flexoelectric effects
(s.\ {}Table~\ref{tab:coefficients}).  A negative value of $\Gamma$ is
required for the occurrence of abnormal rolls (see
Sec.~\ref{sec:homogeneous}) and chevron patterns.

The terms associated with $S_x$ and $S_z$ in Eqs.~(\ref{vx},\ref{vz})
represent internal stresses of the convection pattern.  As for the
coupling of $A$ to $\vec v$ in Eq.~(\ref{full3da}), a corresponding
term for the $y$ direction or a term of the form $S_{yx} \dy {\rm
  Im}\{{A^* \,\dx\, A}\}$ do not enter Eq.~(\ref{vy}), although they
are allowed by symmetry.  $S_{xx}$, $S_{yy}$, \ldots, $S_{zx}$ can be
interpreted as surface tensions of the planes of equal phase of the
convection pattern.

The high number of soft modes and the rich, non-potential coupling
almost certainly lead to spatio-temporally chaotic states already at
onset, provided the spatial extensions of the sample are large enough.

\section{Effectively 2D pattern dynamics near threshold}
\label{sec:dd}

Currently more interesting than the fully 3D chaotic state are, from
the experimental point of view, the quasi 2D pattern dynamics in a
restricted geometry near threshold [$\epsilon^\prime \lesssim
(\lambda/d)^2$].  The dynamics is here derived for a slight
generalization of the usual setup:  A magnetic field $H_y$ might be
applied along the $y$ direction (i.e.\ {}in-plane, normal to the
rubbing direction).  With $H_y$ slightly below the twist \fred{} field
$\chi_a H_F^2=k_{22} (\pi/d)^2$, the amplitude of the twist mode, which
is known to be important for the pattern dynamics in any case, becomes
a slow variable and must be included explicitly in the 2D formalism.
The conventional setup is described by $H_y=0$.

\subsection{Derivation of the 2D description}
\label{sec:dd_derivation}

The appropriate boundary conditions for the soft modes at the
enclosing electrodes are 
\begin{align}
  \label{bc3d}
  \Phi_0,\Phi_r,\Phi_i,\varphi,\vec v=0
  \quad\text{at}\quad
  z=\pm d/2.
\end{align}
Thus, all components of the electric potential and the velocity field
are damped by the boundaries.
    
Again the ``center manifold'' method is used (s.
Appendix~\ref{sec:methods}), now to reduce the 3D equations to 2D.
The dynamically active part $S$ of the state vector $U$ is now given
by the sum of
\begin{align}
  A(x,y,z)&=A^\prime(x,y) \cos(\pi z/d),\\
  \varphi(x,y,z)&=
  \varphi^\prime(x,y) \cos(\pi z/d),\\
  \druck(x,y,z)&=\druck^\prime(x,y),
\end{align}
with the amplitudes of the active modes $A^\prime$,
$\varphi^\prime$, and $\druck^\prime$.
    
To be specific, associate $A$ with the complex conjugate of
Eq.~(\ref{full3da}) and $\Phi_0$, $\Phi_r$, $\Phi_i$, $\varphi$,
$v_x$, $v_y$, $v_z$, and $\druck$ with successive equations in the
system (\ref{mfeqs}), and define the scalar product
$\left<\cdot|\cdot\right>$ as usual as the equally weighted sum over
$z$ integrals over products of the two components.  The projector onto
the slow dynamics is constructed from the bi-orthonormalized linear
functionals $(2/d) \int_{-d/2}^{d/2} \cos(\pi z/d) \,A \, d z$, $(2/d)
\int_{-d/2}^{d/2} \cos(\pi z/d) \,\varphi\, dz$, and $(1/d)
\int_{-d/2}^{d/2} \druck \, dz$.

We proceed with a calculation of the excitations in the fast subspace
$R$ by a term-by-term solution of Eq.~(\ref{nlreductive-dyn}).  The
truncation is chosen such that the distinguished limit
\begin{gather*}
\notag 
  \druck^\prime,H_F^2-H_y^2,\dt \sim {\varepsilon^\prime} ,
  \\
  \begin{align}  
    \label{distinguished_limit}
    \varphi^\prime,A^\prime,\dx,\dy \sim {\varepsilon^\prime}^{1/2} ,
  \end{align}
  \\
  {\varepsilon^\prime}=O (\lambda/d)^{4},\quad\text{and}\quad
  d\to\infty
  \notag
\end{gather*}
is correctly described.

At linear order in the amplitudes and to linear order in $\dx,\dy$,
the slaved part of the state vector contains only the contributions
\begin{align}
  \label{p0flow}
  v_x^{(1)}&=-\frac{1}{\eta_2}
  \left(\frac{d^2}{8}-\frac{z^2}{2}\right)\dx \druck^\prime,\\
  v_y^{(1)}&=-\frac{2}{\alpha_4}
  \left(\frac{d^2}{8}-\frac{z^2}{2}\right)\dy \druck^\prime.
\end{align}
At order $O(|A^\prime|^2)$ and without any $x$ or $y$ modulations
there are excitations of the electric field
\begin{align}
  \label{a2pots}
  \Phi_r^{(2)}+i \Phi_i^{(2)}=&\frac{I_r+i I_i}{\sigma_\perp+i \omega
    \epsilon_\perp} \int_{-d/2}^z 
  |A(z^\prime)|^2-\left<|A|^2\right>_z d z^\prime ,
\end{align}
and a contribution to the pressure field, orthogonal to the active
pressure mode,
\begin{align}
  \label{a2druck}
  \druck^{(2)}=&-\left(S_z-2 S_E
  \right) \cdot
  \left(|A|^2-\left<|A|^2 \right>_z\right),
\end{align}
where $\left<|A|^2 \right>_z=|A^\prime|^2/2$ is the average of $|A|^2$
over $z$ and
\begin{align}
  \label{SE}
  S_E:=\frac{E_0 \epsilon_\perp (\omega
    \epsilon_\perp I_i+\sigma_\perp I_r)}{
    \omega^2 \epsilon_\perp^2+\sigma_\perp^2}.
\end{align}
By gradients of $\Phi_r^{(2)}$, $\Phi_i^{(2)}$, $\druck^{(2)}$, and by
direct contributions at order $O(\dx |A^\prime|^2,\,\dy |A^\prime|^2)$
the mean flow
\begin{align}
  v_x^{(2)}=& 
  \left[S_x
    \left(\frac{d^2}{16}-\frac{z^2}{4}\right)+
  \right.
  \notag\\ &
  \left. \vphantom{\left(\frac{d^2}{16}-\frac{z^2}{4}\right)}
    (S_E+S_x-S_z)\frac{d^2}{4 \pi^2}\cos^2\!\!\left(\frac{\pi
        z}{d}\right) \right] \frac{\dx
    |A^\prime|^2}{\eta_2},
  \\
  v_y^{(2)}=& (S_E-S_z)\frac{d^2}{4
    \pi^2}\cos^2\!\!\left(\frac{\pi z}{d}\right) \frac{2 \dy
    |A^\prime|^2}{\alpha_4}.
\end{align}
is excited (there is no distinction between ``singular'' and
``non-singular'' mean flow in this approach).  Excitations of $\Phi_0$
are of the order $O(\varepsilon^\prime)$ and do, as all other
remaining corrections, not contribute at leading order.
    
Projection of the dynamics with the full state vector $U=S+R$ onto the
slow space yields the equations of motion for the pattern amplitude
\begin{subequations}
  \label{a2d}
  \begin{align}
    \tau\partial_t A^\prime= \Biggl[&\xi_x^2\partial_x^2 +
    \xi_y^2\left(\partial_y^2-
      \frac{16 i q_c}{3 \pi}\varphi^\prime \partial_y-\frac{3q_c^2}{4} {\varphi^\prime}^2
    \right)  
    \\
    & \label{locala2d}
    + i \frac{8}{3 \pi} \beta_{y}
    \nabla_y\varphi^\prime
    +\epsilon' - \left(\frac{3}{4}g+\frac{S_E}{
        \epsilon_\perp E_0^2}\right) |A^\prime|^2 
    \\
    &\label{amfona}
    -\frac{i d^2 q_c \tau}{48 \pi^2}\,
    \frac{9 S_E+(15+2\pi^2) S_x-9 S_z}{\eta_2}
    \nabla_x |A^\prime|^2
    \\ 
    & \label{pmfona}
    +\frac{i d^2 q_c \tau}{12 \pi^2}\,\frac{3+\pi^2}{\eta_2}
    \nabla_x \druck^\prime
    \Biggr] A^\prime
  \end{align}
\end{subequations}
and the twist mode,
\begin{subequations}
  \label{phi2d}
  \begin{align}
    \label{linearphi}
    \gamma_1 \dt \varphi^\prime =& \left(k_{33} \dx^2+k_{11}^\prime
      \dy^2 -\chi_a
      (H_F^2-H_y^2)\right) \varphi^\prime \\
    \label{Gamma_term}
    &+\frac{q_c \Gamma}{2} {\rm Im}\left\{
      {A^\prime}^*\left[
        \frac{8}{3\pi}\dy-\frac{3 i q_c}{4}\varphi^\prime
      \right]A^\prime
    \right\}\\
    \label{pmfonphi}
    & +\frac{4 d^2}{\pi^3}
    \left[\frac{\alpha_3}{\eta_2}+\frac{2
        \alpha_2}{\alpha_4}\right] \dx \dy \druck^\prime \\
    \label{amfonphi}
    &-\frac{2 d^2}{3 \pi^3}\,
    \left[
      \frac{2 \alpha_2(S_E-S_z)}{\alpha_4 }
    \right.
    \notag \\ & \quad
    +
    \left.
      \frac{\alpha_3 (S_E+4 S_x-S_z)}{\eta_2}\right]\,
    \dx\dy|A^\prime|^2, 
  \end{align}
\end{subequations}
and the pressure Poisson equation \cite{croho}
\begin{subequations}
  \label{Peq}
  \begin{align}
    0=&-\frac{d^2}{12}\left[ \frac{1}{\eta_2}
      \dx^2 + \frac{2}{\alpha_4} \dy^2 \right]\druck^\prime\\
    \label{dxaonp}
    &+\frac{d^2}{8 \pi^2}\, \frac{S_E+(1+\pi^2/3)
      S_x-S_z}{\eta_2}\, \dx^2 |A^\prime|^2\\
    \label{dyaonp}
    &+\frac{d^2}{4
      \pi^2}\,\frac{S_E-S_z}{\alpha_4}\dy^2 |A^\prime|^2.
  \end{align}
\end{subequations}

\subsection{Remarks on the 2D amplitude equations}
\label{sec:2ddiscuss}

By the explicit representation of the coupling coefficients in
Eqs.~(\ref{a2d}-\ref{Peq}), the increasing importance of the mean flow
contributions in lines (\ref{amfona}, \ref{pmfona}, \ref{pmfonphi},
\ref{amfonphi}) as the separation $d$ of the damping boundaries
increases becomes obvious.  On the other hand, for small enough $d$,
i.e.\ {}${\varepsilon^\prime}\ll(\lambda/d)^4$, mean flow is
negligible.  With some rescaling (indicated by a caret $\check {\text
  \ }$) dynamics are then described by the system
\begin{subequations}
\begin{align}
  \begin{split}
    \label{scaled-NW-normal-A}
    \check \tau \partial_{\check t} \check A =\Big [& 1+
    \partial_{\check x}^2+\partial_{\check y}^2-2 i c_1 \check
    \varphi\partial_{\check y} - c_2 \check \varphi^2
    \\
    &-|\check A|^2 + i \check \beta_{\check y} \check \varphi_{,y}\;
    \Big ]\, \check A,
  \end{split}\\
  \begin{split}
    \label{scaled-NW-normal-phi}
    \partial_{\check t} \check \varphi =& \partial_{\check y}^2 \check
    \varphi+\check K_3 \partial_{\check x}^2 \check \varphi -
    \check H^2 \check \varphi
    \\
    &+ 2 \check \Gamma \;  
    {\mathrm{Im}}\left\{\check A^* (\partial_{\check y} - i \check
    \varphi) \check A \right\}.
  \end{split}
\end{align}
\end{subequations}
This is the ``normal form'' for the dynamics of a pattern coupled to a
soft mode with symmetry under the reflections $[ \check x \to -\check
x,\,\check \varphi \to -\check \varphi,\, \check A \to \check A^* ]$
and $[ \check y \to -\check y,\, \check \varphi \to -\check \varphi]$
or $[ \check y \to -\check y,\, \check \varphi \to -\check \varphi,\,
\check A \to -\check A ]$.  In the present case $c_1=c_2=\frac{4}{3}
\left(\frac{8}{3 \pi}\right)^2 \approx 0.96$ are fixed by geometric
constraints \cite{linddpl}.  These values are quite close to the case
with an underlying rotation symmetry $c_1=c_2=1$ \cite{rohekrpe},
which turns out to be somewhat singular in its dynamical properties
\cite{krzapriv,roth}.

With $H_y^2=0$ in line (\ref{linearphi}), $\varphi^\prime$ is damped
out and becomes one of many higher order corrections.
Equations~(\ref{a2d}, \ref{Peq}) with $\varphi=0$ are then sufficient,
and with ${\varepsilon^\prime}\ll(\lambda/d)^4$ the simple
Ginzburg-Landau Equation~(\ref{rgle}) with
\begin{align}
  \label{gprime}
  g^\prime=\frac{3}{4}g+\frac{S_E}{\epsilon_\perp E_0^2}
\end{align}
(the second
contribution is numerically small, see Table~\ref{tab:coefficients})
is retained.

When modulations of $|A|^2$ along $x$ are strong, like in the chevron
pattern, it is instructive to redefine the pressure
$\druck^\prime\to\druck^\prime-{\mathrm{const.}}\times |A^\prime|^2$
such that its excitation by $\dx^2
|A^\prime|^2$ in line~(\ref{dxaonp}) is canceled.  It turns
out that the remaining excitation of $\druck^\prime$ by $\dy^2
|A^\prime|^2$ is proportional to $S_x$.  The substitute for
line~(\ref{amfona}) incorporating the correction in $\druck^\prime$
does then have the form
\begin{align*}
      -\frac{i d^2 q_c \tau (\pi^2-6)}{16 \pi^4}\,
    \frac{S_E+S_x-S_z}{\eta_2}
    \nabla_x |A^\prime|^2 
\end{align*}
and accounts for a mean flow with a nontrivial flow profile with zero
$z$ average.

\subsection{Pressure \textit{vs.}\ singular mean flow}

\label{sec:singular_mean_flow}

As for any incompressible, Newtonian fluid, Eq.~(\ref{Peq}) is of the
form
\begin{align}
  \label{Pform}
  0=\partial_{\vec r} \cdot  \mathbf{M}\, \partial_{\vec r}
  \druck^\prime-\partial_{\vec r} \cdot \vec V.
\end{align}
In the present simple case, the matrix $\mathbf{M}$ is
constant and, in the usual coordinates, diagonal.  The inhomogeneity
$\vec V$ (with dimensions of velocity) depends on time, $x$, and $y$.
Equation~(\ref{Pform}) can be solved formally by the transformation
\begin{align}
  \label{pg}
  \partial_{\vec r} \druck^\prime=
  {\mathbf{M}}^{-1}\;
  \left(
    \hat z \times \partial_{\vec r} G+\vec V
  \right),
\end{align}
which requires a fundamental solution $G(x,y,t)$ satisfying
\begin{align}
  \label{Gform}
  0=&\hat z \times \partial_{\vec r}\cdot \partial_{\vec r}
  \druck^\prime
  \notag\\
  =&(\hat z \times \partial_{\vec r}) {\mathbf{M}}^{-1}(\hat z \times
  \partial_{\vec r}) G+(\hat z \times \partial_{\vec
    r}){\mathbf{M}}^{-1}\vec V .
\end{align}
With some rearrangements, Eq.~(\ref{Gform}) has the same simple
structure as Eq.~(\ref{Pform}).  The quantity $G$ can be interpreted
as a stream function generating a certain component of the large-scale
variations of the velocity field $(v_x,v_y)\sim(\hat z \times
\partial_{\vec r})G$, the ``singular mean flow'' (in early works
\cite{szmf} expressed in terms of the vertical vorticity $-\nabla^2
G$).  By using Eq.~(\ref{Gform}) instead of Eq.~(\ref{Pform}) and
eliminating the pressure $\druck^\prime$ \textit{via} Eqs.~(\ref{pg})
also in the remaining equations [in our case (\ref{a2d},\ref{phi2d})],
a description fully in terms of the singular mean flow is obtained.

In principle the three forms (\ref{Pform},\ref{pg},\ref{Gform}) are
equivalent, although the last is often preferred in the literature
(for an overview see both Refs.~\cite{croho,nps}), perhaps because in
some situations with high symmetry $G$ is not excited, while
$\druck^\prime$ is.  When disregarding the effects of the additional
soft mode $\varphi^\prime$, application of the  less obvious but direct
method of Kaiser and Pesch \cite{kp} leads to the same result for the
mean flow equation~(\ref{Gform}) as the route described here.

The flux $\vec V$ in Eq.~(\ref{pg}) is determined by
Eq.~(\ref{Pform}) only up to a transformation $\vec V \to \vec V +
(\hat z \times \partial_{\vec r}) G_0(x,y,t)$ which implies a
redefinition $G \to G - G_0$.  Hence $(\hat z \times \partial_{\vec
  r}) G$ is not generally proportional to the $z$ average of the large
scale flow.  This is not necessary for the formalism to work.  In
simple cases, like the present, the stream function $G$ has this
property with the ``natural'' choice of $\vec V$.  To guarantee it in
general, the method of Newell, Passot and Souli \cite{nps} can be used
to derive Eq.~(\ref{pg}) (Eq.~(2.59) in \cite{nps})
directly under this additional constrain.

Here, following Ref.~\cite{croho}, the formulation as a mass
conservation equation~(\ref{Pform}) is used, since it derives
naturally from the general formalism and is thus easily extended to
three spatial dimension, to additional homogeneous soft modes, or to
relax the assumption of incompressibility (which is not essential for
the relevance of mean flow as is sometimes suggested).

In principle, terms proportional to $\dx^2 \varphi^\prime$, $\dy^2
\varphi^\prime$ could also appear in the $\druck^\prime$
equation~(\ref{Peq}).  It is a particularity of the system considered
here that they do not.  In the conduction regime of EC in cells with
homeotropic boundary conditions it can be shown \cite{roth} that there
is such a, presumably small, excitation of $\druck^\prime$ by
$\varphi^\prime$ proportional to the dynamic flexoelectric coefficient
$\zeta^E$.

\subsection{Variation of the  boundary conditions}
\label{sec:other_BC}

Of course, the reduction from 3D to 2D can also be carried out for
other boundary conditions than (\ref{zboundary},\ref{bc3d}).  Some
variants are of practical interest.

An effect similar to applying a magnetic field along $y$ can be
obtained by a homeotropic (normal) anchoring of the director at the
boundaries: For negative, not too small $\epsilon_a$, and with
external electric fields as required for dielectric EC, a homeotropic
director alignment is unstable and instead the director becomes
planarly oriented everywhere, except for small boundary layers of
thickness $(k_{11}/\epsilon_a)^{1/2} E_0^{-1}\approx\lambda$.  Hence,
on the length scales $\gg \lambda$ relevant for the amplitude
formalism, the boundary layer vanishes and instead free boundary
conditions for $\varphi$ can be assumed.  For this setup only some
geometry factors have to be changed in Eqs.~(\ref{full3da}-\ref{Peq}).

Another variant is a twisted cell, e.g.\ {}with $\varphi=0$ at
$z=-d/2$ and $\varphi=\pi/2$ at $z=d/2$, as it was recently
investigated experimentally by Bohatsch and Stannarius \cite{bosta}.
Based on the reduced 3D description derived here, the linear theory
for this configuration is developed in Ref.~\cite{twisted}.

\subsection{Higher order contributions}
\label{sec:higher_order_terms}

It is worth noticing that, when only the limit of small
$\varepsilon^\prime$ is considered while keeping $d$ fixed, there are
several other nonlinear and higher order gradient terms besides those
in lines~(\ref{amfona},\ref{pmfona}) which are formally of the same
order of magnitude.  A longer list containing more than fifty terms
has been calculated numerically by Kaiser and Pesch \cite{kp} for the
conduction regime of EC.  Their form correctly predicts the stability
of ideal roll patterns close to threshold but, since it does not
separate mean flow and director effects, is unable to describe
important effects like the transition to abnormal rolls \cite{pdrkp}
(see Sec.~\ref{sec:homogeneous}).  These limitations are partly overcome
in the less systematic but numerically surprisingly accurate
description of Plaut and Pesch \cite{plpe}.

In fact, since the lowest order mean flow effects entering
Eqs.~(\ref{a2d}-\ref{Peq}) all depend on gradients of $|A^\prime|^2$,
they do not contribute to long-wavelength instabilities of the band
center (i.e.\ {}$A\equiv {\mathrm{const.}}$).  For the calculation of
the thresholds of long-wavelength instabilities of the pattern it may
be useful to formally set up amplitude equations including higher
order mean flow.  But for a systematic quantitative description of
{\em general\/} non-ideal patterns containing structures of size $\sim
\lambda \epsilon^{-1/2}$ only the truncation used here is justified.  When
higher order mean flow becomes relevant the 2D amplitude formalism is
already breaking down because, for example, the coherence length
$\xi_y\varepsilon^{-1/2}$ becomes of the order of the sample thickness
\cite{kp}.

\section{Stability of ideal patterns}
\label{sec:stability}

Rather than deriving stability bounds of ideal patterns $A=a(z) \exp(i
q x +i p y)$ using a reduced 2D description, it seems more appropriate
to do the calculations directly based on the 3D equations, in
particular when $H_y=0$.  This is easily seen from the fact that there
is a (numerically small) manifest deviation of $a(z)$ from the
$\cos(\pi z/d)$ profile at the threshold of all instabilities
calculated below, indicating that the expansion for small $A^\prime$
is breaking down.  Nevertheless, in order to obtain analytic
estimates, Galerkin approximations will be used which correspond
effectively to Eqs.~(\ref{a2d}-\ref{Peq}) and their extension to
higher order contributions.  For simplicity, the stability analysis
shall here be performed only at the band center $q,p=0$ and only take
homogeneous and long-wavelength instabilities with modulations along
$y$ into account.  The latter restriction is justified by experimental
observations and by the fact that, with this geometry, the advection
of the patter by mean flow is particularly strong.  It is then
sufficient to consider only the interaction of $\varphi$, $v_x$, and
of the phase $\theta$ given by $A=a(z) \exp i \theta(y,z,t)$
[$a=a(z)=O(\varepsilon^\prime/g^\prime)^{1/2}$ be real and given by
Eqs.~(\ref{full3da},\ref{fir},\ref{fii})], which leads to the
following linear problem:
\begin{subequations}
  \label{stabEqs}
  \begin{align}
    \begin{split}
      \label{nyEq}
      {{\gamma}_1}\,
      {{\partial}_t} \,{\varphi}
      =
      &\left[   
        \left(
          {k_{{11}}}+\smash[b]{{2\, \nen\, e_-^2/\epsilon_a}}
        \right)
        \,{\partial_y^2} + 
        {k_{{22}}}\,{\partial_z^2}
      \right] \,
      {\varphi}
      \\
      &- 
      {{\alpha}_3}\,{{\partial}_y}\,{v_x} 
      +
      (q_c { \Gamma}/2)\,a^2\, ({{\partial}_y} \theta- q_c \varphi)
      ,
    \end{split}
    \\
    \begin{split}
      \label{vxEq} 
      \rho_m\,{{\partial}_t}\, {v_x} = & \, \eta_2\, \left(
        {\partial_y^2} + {\partial_z^2} \right) \, {v_x}
%       \\
%       &
      +
      {{\alpha}_3}\,{{\partial}_y}\,{{\partial}_t}\,{\varphi}\phantom{.}
      \\
      &+ S_{yy}\, a^2 \,{{\partial}_y} ({{\partial}_y} \theta - q_c
      \varphi) 
      \\&
%       + S_{zz}\, a^2\, \left[2\, (\nabla_z \ln a)\, \dz\, \theta +
%         \dz^2\,\theta\right]
      + S_{zz}\, \dz \left( a^2 \, \dz\, \theta \right)
      ,
    \end{split}
    \\
    \begin{split}
      \label{thetaEq}
      \tau \dt \theta =&\,
      -\tau q_c v_x 
      + \beta_y \dy \varphi
      + \xi_y^2 \dy^2 \theta
      \\ &
%       + \xi_z^2 \left[2\, (\nabla_z \ln a)\, \dz\, \theta +
%         \dz^2\,\theta\right]
      + \xi_z^2\,a^{-2}\, \dz \left( a^2 \, \dz\, \theta \right)
      .
    \end{split}
  \end{align}
\end{subequations}
Since $a=0$ at $z=\pm d/2$, the singular last term in
Eq.~(\ref{thetaEq}) implies boundary conditions $\dz \theta=0$.  The
other boundary conditions are $\varphi, v_x=0$ at $z=\pm d/2$.

\subsection{Homogeneous destabilization}
\label{sec:homogeneous}

First, consider homogeneous ($\dy=0$) destabilizations of the pattern.
In this case $\varphi$ decouples from $v_x$ and $\theta$.  The
destabilization of $\varphi$ is known as the abnormal-roll instability
and was investigated in the dielectric regime in Ref.~\cite{linddpl}.
It was found that
\begin{align}
  \label{epsilonAR}
%epsAN = (-8*gs*k33*Pi^2)/(3*d^2*Gamma*qc^2)
  \varepsilon^\prime=\varepsilon_{AR}^\prime:=-\frac{8\, \pi^2\,
    g^\prime\, k_{33}} {3\, d^2\, q_c^2\, \Gamma},
\end{align}
which can be derived from Eqs.~(\ref{a2d},\ref{phi2d}), is typically a
good approximation of the threshold.  In MBBA
$\varepsilon_{AR}^\prime\approx 2.8\times (\lambda/d)^2$
($\omega\tau_0\ge 8$).  The value obtained by numerically calculating
eigenmodes of Eq.~(\ref{nyEq}) directly is $3\%$ lower than the value
of Eq.~(\ref{epsilonAR}).  Below it will be shown that for MBBA the
abnormal-roll instability is precede by a long-wavelength modulation
instability.  Nevertheless, some phenomena associated with abnormal
rolls might be observable around $\varepsilon=\varepsilon_{AR}$, e.g.\ 
{}the tendency of defects in the convection pattern to cluster along
lines parallel to the rolls.

For the discussion of homogenous perturbations of $v_x$ and $\theta$
notice first the neutral mode associated with a translation of the
pattern $\theta\to\theta+\text{const.}$.  This mode is best dealt with
by decomposing $\theta$ as $\theta=\tilde \theta(z,t)+\Theta(t)$ such
that the $z$ average $\left<\right.\! \tilde \theta \!\left.\right>_z$
vanishes.  By multiplying Eq.~(\ref{thetaEq}) by $a^2$ and integrating
over $z$ the dynamics of $\Theta$ is obtained as
\begin{align}
  \label{ThetaEq}
  \dt \Theta= -q_c\frac{\left<a^2 v_x\right>_z}{\left<a^2\right>_z}.
\end{align}
This describes the advection of the pattern by mean flow at large
pattern amplitudes.  

At the threshold of instability one has $\dt \tilde\theta,\dt v_x=0$
and $\dt\Theta=\text{const.}$ ($\dt\Theta\ne0$ implies an acceleration
instability of pattern and liquid crystal).  To calculate the
threshold, eliminate $\tilde\theta$ from Eq.~(\ref{vxEq}) by
Eqs.~(\ref{thetaEq},\ref{ThetaEq}), obtaining the equation
\begin{align}
  \label{driftCriterium}
  0=\eta_2\dz^2 v_x+
  \frac{S_{zz} \tau q_c}{\xi_z^2}a^2
  \left[v_x-\frac{\left<a^2 v_x\right>_z}{\left<a^2\right>_z}\right]
\end{align}
from which the critical mode can be determined numerically.  When
using the low-amplitude approximation
$a^2=(\varepsilon^\prime/g^\prime)\cos^2(\pi z/d)$ the critical mode
is found to be antisymmetric in $z$ (i.e.\ {}$\dt \Theta=0$) at
\begin{align}
  \label{lowDriftThreshold}
  \varepsilon^\prime=\varepsilon^\prime_{\text{drift}}=
  73.3\, \frac{\eta_2\, g^\prime\, \xi_z^2}{d^2\, S_{zz}\, \tau\, q_c}.
\end{align}
[With the large amplitude approximation $a^2=\varepsilon/g$, the first
symmetric and antisymmetric mode both become unstable at the
\emph{same} $\varepsilon=(4 \pi^2\eta_2 g \xi_z^2)/(d^2 S_{zz} \tau
q_c)$.]  The antisymmetric excitation of the phase $\tilde\theta$
involved in this instability is obviously inaccessible to a reduced 2D
description.  Since $S_{zz}<0$ in MBBA, the value of
$\varepsilon^\prime_{\text{drift}}\approx -100 \times (\omega
\tau_0)^{-1} (\lambda/d)^2$ is negative and the instability does not
occur.  However, since the electric contribution to $S_{zz}$ [the
second term in formula~(\ref{Szz})] is always positive and comparable
in size with the hydrodynamic one, a positive $S_{zz}$ is thinkable
for other materials.  When $S_{zz}<0$, the mechanism leading to
Eq.~(\ref{driftCriterium}) is stabilizing -- in particular for all
perturbations of $\tilde\theta$: The lamellae of EC are forced to
align normal to the boundaries.  This explains why the experimental
shadowgraph images, which average the pattern along $z$, remain quite
sharp even for complicated pattern dynamics.

\subsection{Modulation instabilities}
\label{sec:modulation}

The stability of ideal patterns with respect to perturbations in
$\varphi$, $v_x$, and $\theta$ modulated with small wave numbers $k$
along $y$ was investigated numerically using the
system~(\ref{stabEqs}).  It was found that for MBBA the dominating
destabilizing feedback loop is based on the excitation of $v_x$ by the
term containing $S_{yy}$ in Eq.~(\ref{vxEq}) and advection of the
phase.  There is a good Galerkin approximation for the numerical
results.  Using the low amplitude approximation for $a^2$, Galerkin
modes $\varphi,v_x\sim\cos(\pi z/d)$ and $\theta\sim 1$, and
projectors $\int\cos(\pi z/d) \cdot \dz$ on
Eqs.~(\ref{nyEq},\ref{vxEq}) and $\int\cos^2(\pi z/d) \cdot \dz$ on
Eq.~(\ref{thetaEq}) to reduce the system~(\ref{stabEqs}) to algebraic
equations, the threshold for long-wavelength modulations instabilities
is estimated to be at
\begin{align}
  \begin{split}
  \label{epsilonZZ}
%   epsLongZZ = (72*gs*k33*nu2*Pi^4*xiyy2)/
%   (d^2*qc*(-256*betay*Gamma*nu2 + 512*k33*Syy*tau - 
%   27*Gamma*nu2*Pi^2*qc*xiyy2))
  \varepsilon^\prime&=\varepsilon^\prime_{ZZ}:= \frac{72 \pi^4 \eta_2
    g^\prime k_{33} \xi_{y}^2} {d^2 q_c}\times 
  \\  &\left[ 512
    k_{33} S_{yy}\tau -\eta_2 \Gamma \left(256 \beta_y+27 \pi^2 q_c
      \xi_{y}^2 \right)\right]^{-1}.
\end{split}
\end{align}
With $\beta_y=-q_c\xi_{yy}^2$ the parenthesis following $\Gamma$
nearly vanish; the remainder is related to the small deviation of
$c_1$ in Eq.~(\ref{scaled-NW-normal-A}) from unity.  Numerically
$\varepsilon^\prime_{ZZ}\approx 9 \times (\omega \tau_0)^{-1}
(\lambda/d)^2$ (observe that $\varepsilon^\prime_{ZZ}$ and
$\varepsilon^\prime_{AR}$ have different frequency dependence).  In
particular, at $\omega \tau_0=8$, using the second lowest Fourier
approximation without flexoelectric effects, Eq.~(\ref{epsilonZZ})
yields $\varepsilon^\prime_{ZZ}=0.792 (\lambda/d)^2$ while the
numerical solution of system~(\ref{stabEqs}) gives a threshold at
$\varepsilon^\prime=0.797 (\lambda/d)^2$.  Using the Galerkin
approximation it is easily verified that the instability is in fact of
the long-wavelength ($k\to 0$) type.

\section{Qualitative transitions}
\label{sec:transitions}

For the case $H_F=0$, order of magnitude estimates shall be used to
distinguish regions of qualitatively different pattern dynamics in
parameter space - above as well as below the stability bounds of ideal
patterns.

As mentioned before, simple Ginzburg Landau dynamics can be expected
for small $\epsilon$ until the last two lines in Eq.~(\ref{a2d})
become relevant.  Assuming $\dx \approx
{\varepsilon^\prime}^{1/2}\xi_x^{-1}$ and $|A^\prime|^2\approx
\varepsilon^\prime/g^\prime$ , these terms have an effect of the
magnitude of $\varepsilon^\prime A^\prime$ when, say,
$\varepsilon^\prime\approx [3\,\pi^2\, \eta_2\, g\, q_c\, \xi_x\,
\lambda^2]^2/[(\pi^2-6)\, (S_E+S_x-S_z)\,\tau\, d^2]^2\approx 2.\times
10^4 \times (\lambda/d)^4 (\omega\tau_0)^{-2}$ (MBBA,
$\omega\tau_0\gtrsim 10$).  Preliminary simulations of
Eqs.~(\ref{a2d},\ref{Peq}) with $\varphi^\prime\equiv0$ show that for
higher $\varepsilon^\prime$ defect cores (where $\dx |A^\prime|^2$ is large)
are strongly deformed and lines along which the phase $\arg A^\prime$
``jumps'' are often generated and long living (rather long living
phase jump lines are also observed experimentally and in simulations
of a similar model \cite{kp}; it is not clear, though, whether these
are due to lowest order mean flow effects).  But the simulations also
indicate that these lowest order mean flow effects, although they are
formally dominating over the direct nonlinear saturation \textit{via}
the last term in line~(\ref{locala2d}), do not prevent the system form
finally reaching a steady state with $|A^\prime|^2\approx
\varepsilon^\prime/g^\prime$.

Thus, assuming still $|A^\prime|^2\approx\varepsilon^\prime/g^\prime$, there
will be a further transition at $\varepsilon^\prime\approx[3^{1/2} \pi\,
(\alpha_4+2 \eta_2) g k_{33} q_c^3 \xi_y/(128 \alpha_2
S_x)]^{2/3}\times(\lambda/d)^{8/3}\approx 1.\times(\lambda/d)^{8/3}$
(MBBA, $\omega\tau_0\ge 8$) where contributions from slaved excitations
of the in-plane director $\varphi$ by mean flow become relevant in the
2D pattern dynamics.  Semiquantitatively these effects are described by
Eq.~(\ref{phi2d}) with $H_y=0$, but a restriction of dynamics to a
single linear mode of $\varphi$ is then not justified.

When, with increasing $\epsilon$, horizontal length scales become of
the size of the sample thickness $d$ [at $\varepsilon^\prime\approx
(\pi \xi_x/d)^2\approx 0.2 \times (\lambda/d)^2$ (MBBA), say] the 2D
description breaks down.  Because then $d$ is not the dominating
length scale for the damping of mean flow anymore, the trend in the
influence of mean flow on the smallest structures in the $A$ field
(e.g.~defects) is reversed.  Now the structures themselves set the
length scale.  Assuming $\dx \approx \varepsilon^{1/2}\xi_x^{-1}$ and
$|A|^2\approx \varepsilon/g$, the contribution $i \tau v_x A$ in
Eq.~(\ref{full3da}) is large compared to $\varepsilon A$ up to
$\varepsilon\approx(q_c S_x \tau \xi_x/\eta_2 g)^2\approx 1.\times
10^{-6}\times (\omega\tau_0)^2 $ (MBBA, $\omega\tau_0\gtrsim 10$).
For larger $\varepsilon$ the cores of defects are not affected by
lowest order mean flow effects.  Larger structures, like the phase
field of the pattern or variations in the defect density (e.g.\ {}in
chevron patterns), may still be.

Current experimental resolutions are of the order $\Delta
\varepsilon=10^{-3} \cdots 10^{-2}$.  The estimates above suggest that
lowest order mean flow effects are best observed near the upper limit
of the frequency range for the validity of the hydrodynamic
description used here, at $\omega\tau_0 = O(10^2)$ (see
Section~\ref{sec:dimensional}).  With $d \approx 5 \lambda$ they
should be observable in the range of validity of the 2D description.

According to the model for the chevron mechanism~\cite{RoKrChev},
chevrons depend essentially on the abnormal-roll mechanism and can only
form above the abnormal-roll instability bound [here given by
Eq.~(\ref{epsilonAR})].  Thus, it is plausible to assume a
$\varepsilon^\prime\sim(\lambda/d)^2$ threshold for chevron formation,
in accordance with measurements presented in Ref.~\cite{scheukp},
where an approximate $\omega^{-1}$ frequency dependence is found.

% The importance of the term~(\ref{Gamma_term}) in the model
% \cite{RoKrChev} for the chevron mechanism and also the measured
% approximate $\omega^{-1}$ frequency dependence of the threshold of
% chevron formation \cite{scheukp} suggest that this threshold is tied
% to the transition at $\varepsilon^\prime\approx(\lambda/d)^2$. 

For very high $\varepsilon^\prime$ many authors report the formation
of disclination loops, which, being singularities in the director
field, indicate already a breakdown of \emph{hydrodynamics}.  In
thicker cells, the chevron pattern might decay along other routes
before the 3D amplitude formalism breaks down at some
$\varepsilon=O(1)$.

For convection in most quasi-2D systems, i.e.\ {}systems with
$d/\lambda=O(1)$, all these transitions, from the breakdown of
simple 2D Ginzburg-Landau dynamics to the breakdown of the amplitude
formalism, do, in principle, collapse at $\varepsilon=O(1)$.  Only by
specially designed experiments (as in \cite{rebk}) can these
transitions be unfolded.

\section{Electric Nusselt numbers}
\label{sec:nusselt_numbers}

Recently ``electric Nusselt numbers'' ${\mathcal{N}}_r$, ${\mathcal{N}}_i$
have been introduce by Gleeson, Gheorghiu, Plaut \cite{glghpl} as the
ratio of the in-phase or, respectively, out-of-phase components of the
electric current to the corresponding values expected for the
unstructured basic state at a given voltage, minus one.  As for the
Nusselt number in thermal convection, they are to first approximation
proportional to $|A|^2$.

Several interesting questions can be addressed by measuring electric
Nusselt numbers.  First, it follows from the discussion above that, in
dielectric EC, $|A|^2\approx\varepsilon/g$ for thick enough
cells even far inside the three-dimensionally chaotic range (the
average of $|A|^2$ across the pattern is typically only weakly reduced
in the presence of defects; see, e.g., Ref.~\cite{RoKrChev}).
Measurements of the electric Nusselt numbers therefore seem to be an
effective method to test an essential feature of the theory.  Second,
Nusselt number measurements may also help to identify the qualitative
changes in the dynamics predicted in Section~\ref{sec:transitions}, in
particular since they do, in contrast to optical methods, not loose
their sensitivity in thick cells or with small wavenumbers.  Finally,
the frequency dependence of the Landau coefficient $g^\prime$, which
enters the Nusselt numbers near threshold in a simple way, provides
information on the strength of the dynamic flexoelectric effect (see
Appendix~\ref{sec:coefficients}).  From the derivation of $g^\prime$
it is clear that boundary effects do not interfere in these
measurements.  To obtain the theoretical value for the Nusselt
numbers, take the $x$ and $y$ average (symbol
$\left<\cdot\right>_{xy}$) of Eq.~(\ref{fir}) plus $i$ times
Eq.~(\ref{fii}), which leads to
\begin{multline}
  \label{av_fi}
  (\sigma_\perp+i\omega\epsilon_\perp)\,
  \dz^2\left<\Phi_r+i\Phi_i\right>_{xy}=
  \\
  (I_r+i
  I_i)\,\dz\left<|A|^2\right>_{xy}
  \\
  +(I_{rz}+iI_{iz})\,\dz\left<{\mathrm{Im}}\{A^*\,\dx\,A\}\right>_{xy}.
\end{multline}
Taking the boundary conditions~(\ref{zboundary},\ref{bc3d}) into
account, this implies, similar as for $\Phi_r^{(2)}+i\Phi_i^{(2)}$ in
Eq.~(\ref{a2pots}),
\begin{multline}
  \label{av_currents}
  \frac{1}{2}j_z:=-(\sigma_\perp+i\omega\epsilon_\perp)\,\dz
  \left<\Phi_r+i\Phi_i\right>_{xy}=\\
  (I_r+i
  I_i)\,\left<|A|^2\right>_{xyz}
  \\
  +(I_{rz}+iI_{iz})\,\left<{\mathrm{Im}}\{A^*\,\dx\,A\}\right>_{xyz}
\end{multline}
at $z=\pm d/2$.  Obviously, $j_z$ is the complex amplitude of the
average, pattern-induced electric current density through the sample.
(The last term on the r.h.s.\ {}represents a correction due to a
global deviation of the average wavenumber from the critical one, and
will be dropped below.)  Thus,
\begin{subequations}
  \begin{align}
    \label{Nr}
    {\mathcal{N}}_r=\frac{2 I_r\,\left<|A|^2\right>_{xyz}}{E_0\sigma_\perp}\\
    \intertext{and} 
    {\mathcal{N}}_i=\frac{2
      I_i\,\left<|A|^2\right>_{xyz}}{E_0\epsilon_\perp \omega}.
  \end{align}
\end{subequations}
At threshold,
$(d/d\varepsilon^\prime)\left<|A|^2\right>_{xyz}=(2g^\prime)^{-1}$,
e.g.\ {}$d{\mathcal{N}}_r/d\varepsilon^\prime=0.31$ in MBBA at
$\omega\tau_0=8$, dropping flexoelectric contributions.  The value
increases roughly $\sim \omega$ as frequency increases.  Typically, it
seems to be larger than the corresponding value for the conduction
regime \cite{glghpl}.

\section{Conclusion}
\label{sec:conclusion}

It has been shown how the 3D dielectric convection pattern interacts
with various homogeneous soft modes, which are related to undamped
hydrodynamic modes.  The method to establish these relations is not
unique, but the ``center manifold'' method seems to be favorable over
the ``order parameter'' method.

The reduction of the 3D pattern dynamics to a quasi-2D form in the
layer geometry was derived analytically, thus establishing a
description of the interaction of the pattern with the twist mode and
the pressure field (or singular mean flow).  Scaling analysis suggests
that the transition from the simple, quasi-2D Ginzburg-Landau dynamics
to manifestly 3D dynamics in thick layers unfolds into several well
distinguished steps, the first of which occurs already very close to
threshold ($\epsilon^\prime=O(\lambda/d)^4$).  These characteristics
should generally be expected for 3D patterns.

Ideal, dielectric EC patterns are found to destabilize at some value
$\varepsilon^\prime\sim(\omega\tau_0)^{-1}(\lambda/d)^2$ for which an
analytic approximation in terms of material parameters is given.  A
particular nonlinear mechanism that stabilizes the phase of the
pattern to be constant along $z$, thus giving the pattern a 2D
appearance also at higher $\varepsilon^\prime$, is identified in
Sec.~\ref{sec:stability}.
As outlined in Sec.~\ref{sec:nusselt_numbers}, measurements of the
electric Nusselt numbers are suitable for quantitatively testing the
theory, probing the dynamic flexoelectric effect in nematic liquid
crystals independent of boundary effects, and investigating the route
of the transition from simple 2D to fully 3D dynamics.

It is my pleasure to thank L.~Kramer, Y.~Kuramoto, A.~Lindner,
W.~Pesch, and E.~Plaut for valuable discussion, W.~Decker and W.~Pesch
for providing the basic nematodynamic equations in a computer readable
form, the Kyoto University for its hospitality and the Japan
Foundation for the Promotion of Science (P98285) and the Ministry of
Education, Science, Sports and Culture in Japan for their
support.

\end{multicols}

%\widetext

\begin{widetext}
\begin{subequations}
  \label{mfeqs}
  \begin{align}
    \begin{split}
      \label{fi0}
      0=
      &-
      \left[
        \left(
          {{\sigma}_{\parallel}}- \underline{2\,\nen\,{{\sigma}_a}}
        \right)
        {{\partial}_x^2}\,  
        +
        {{\sigma}_{\perp}}
        \left( 
          {{\partial}_y^2} + {{\partial}_z^2} 
        \right)
      \right]
      \Phi_0\\
      &-
      \underline{2\,\nen\,{e_-}\,({\sigma}_a/
      {\epsilon}_a)\,{{\partial}_x}\,{{\partial}_y}\,\varphi}
    \end{split}
    \\
    \begin{split}
      \label{fir}
      0=&+ 
      {E_{{0}}}
      \,{\partial_x^2}\,
      \left[
        {{\epsilon}_{{\parallel}}}\,
        \omega\,{{\Phi}_i} -
        ({{\sigma}_{\parallel}}- \underline{\sigma_a}) \,{{\Phi}_r} 
      \right] 
      \\
      &+
      {E_{{0}}}\,
      \left( 
        {\partial_y^2} + {\partial_z^2} 
      \right) \,
      \left( 
        {{\epsilon}_{\perp}}\,\omega\,{{\Phi}_i} - 
        {{\sigma}_{\perp}}\,{{\Phi}_r} 
      \right)
      \\
      &-
      \left[ 
        \underline{{{\sigma}_a} \smash[b]{/}{{\epsilon}_a}}
        +
        \left( 
          1 - \underline{3\,\nen} 
        \right) \,
        {{{E_0^2}}}\,
        {{{{\epsilon}_a}}}/
        2\,{{\gamma}_1}
      \right] \,
      \left( 
        {\smash[b]{\underline{{e_+}}}
          - 
          2 {{\gamma}_1}}\;{{\zeta}^E} 
      \right)
      \,
      {{\partial}_x^2}\,{{\partial}_z}\,
      \Phi_0
      \\ 
      &+
      \left[
        \smash[b]{\underline{
            ({\sigma}_a/{\epsilon}_a)
            \left(
              -{k_{{11}}}+{k_{{22}}}
            \right)
            }}
        \vphantom{E_0^2\left(\zeta_E\right)}
      \right.
      \\
      &\quad\quad 
      \left. 
        +
        {e_-}\,{E_0^2}\,
        \left( 
          1 - \smash[b]{\underline{3\,\nen \vphantom{/}}} 
        \right) \,
        \left( 
          \smash[b]{\underline{{{e_+}/2 {{\gamma}_1}}}}
          - 
          {{\zeta}^E} 
        \right)
      \right]
      {{\partial}_x}\,{{\partial}_y}\,{{\partial}_z}\,{\varphi}
      \\
      &+
      \smash[b]{\underline{
          \left({{\sigma}_a}/{  {{\epsilon}_a}}\right)   
          {{{{\partial}_x}\,
              \left({{\alpha}_3}\,{{\partial}_z}\,{v_x}+
                {{\alpha}_2}\,{{\partial}_x}\,{v_z}\right)}}
          }}
      \\
      &+
      {E_{{0}}}\,
      \left( 
        I_r \,{{\partial}_z}\,|A|^2
        +I_{rx} \, \dx\,{\mathrm{Im}}\{A^*\,\dz\,A\}
        +I_{rz} \, \dz\,{\mathrm{Im}}\{A^*\,\dx\,A\}
      \right)
    \end{split}
    \\
    \begin{split}
      \label{fii}
      0=&- 
      {E_{{0}}}
      \,{\partial_x^2}\,
      \left[
        {{{\sigma}_{\parallel}}}\, {{\Phi}_i} +
        ({{\epsilon}_{\parallel}}-\smash[b]{\underline{\epsilon_a}})
        \,\omega\,{{\Phi}_r}
      \right] 
      \\
      &-
      {E_{{0}}}\,
      \left( 
        {\partial_y^2} + {\partial_z^2} 
      \right) \,
      \left( 
        {{\sigma}_{\perp}}\,{{\Phi}_i} + 
        {{\epsilon}_{\perp}}\,\omega\,{{\Phi}_r} 
      \right)
      \\
      &-
      \smash[b]{\underline{\omega\,
          \left( 
            1 - 2\,\nen 
          \right) \,
          \smash[b]{\left( 
              {e_+} - 2\,{{\gamma}_1}\,{{\zeta}^E} 
            \right)} 
          \,
          {{\partial}_x^2}\,{{\partial}_z}\,
          \Phi_0
          }}
      \\ 
      &+
      \smash[b]{\underline{
          \left[
            \omega
            \left(
              -{k_{{11}}}+{k_{{22}}}
            \right)
            -
            (2\, \nen\,e_-/\epsilon_a)
            \,
            \left( 
              {e_+} - 2\,{{\gamma}_1}\,{{\zeta}^E} 
            \right) 
          \right]
          {{\partial}_x}\,{{\partial}_y}\,{{\partial}_z}\,{\varphi}
          }}
      \\
      &+
      \smash[b]{\underline{
          \omega\,
          {{{{\partial}_x}\,
              \left({{\alpha}_3}\,{{\partial}_z}\,{v_x}+
                {{\alpha}_2}\,{{\partial}_x}\,{v_z}\right)}}
          }}
      \\
      &+
      {E_{{0}}}\,
      \left( 
        I_i \,{{\partial}_z}\,|A|^2
        +I_{ix} \, \dx\,{\mathrm{Im}}\{A^*\,\dz\,A\}
        +I_{iz} \, \dz\,{\mathrm{Im}}\{A^*\,\dx\,A\}
      \right)
    \end{split}
    \\
    \begin{split}
      \label{ny}
      {{\gamma}_1}\,
      {{\partial}_t} \,{\varphi}
      =
      &\left[   
        {k_{{33}}}\,{\partial_x^2} + 
        \left(
          {k_{{11}}}+\smash[b]{\underline{2\, \nen\, e_-^2/\epsilon_a}}
        \right)
        \,{\partial_y^2} + 
        {k_{{22}}}\,{\partial_z^2} + 
        {\chi_a H_y^2} 
      \right] \,
      {\varphi}
      \\
      &+
      \left(
        {e_+} - 2\,{{\gamma}_1}\,{{\zeta}^E} 
        -\smash[b]{\underline{2\,\nen\,{e_-}}}
      \right) 
      {{\partial}_x}\,{{\partial}_y}\,
      \Phi_0
      \\
      &- 
      {{\alpha}_3}\,{{\partial}_y}\,{v_x} 
      - 
      {{\alpha}_2}\,{{\partial}_x}\,{v_y}
      +
      {(q_c { \Gamma}/2){\rm Im}\{A^*({{\partial}_y}-i q_c \varphi)A\}}
    \end{split}
    \\
    \begin{split}
      \label{vx} 
      \rho_m\,{{\partial}_t}\,   {v_x} =
      &\left( 
        {{\alpha}_1} + {{\alpha}_4} + {{\alpha}_5} + {{\alpha}_6}
      \right) 
      \,{\partial_x^2}\,{v_x}
      +
      \eta_2\,
      \left( 
        {\partial_y^2} + {\partial_z^2} 
      \right) \,
      {v_x} 
      \\
      &+
      \left( 
        {{\alpha}_2} + {{\eta}_1} 
      \right) \, 
      \dx
      \left( 
        {{\partial}_y}\,{v_y} + \,{{\partial}_z}\,{v_z}
      \right) 
      \\
      &+ 
      {{\partial}_x}\,
      \left[
        -\druck- 
        {E_{{0}}}\,
        {{\epsilon}_{\perp}}\,
        {{\partial}_z}\,{{{\Phi}_r}}  + 
        2\,\alpha_3\,\zeta^E\,
        \left(
          \dy^2+\dz^2
        \right)\,
        \Phi_0
      \right] 
      \\
      &+    
      {{\alpha}_3}\,{{\partial}_y}\,{{\partial}_t}\,{\varphi}\phantom{.}
      {+ S_x {{\partial}_x} |A|^2}
      +
      S_{xx} \dx {\rm Im}\{{A^* \,\dx\, A}\}       \\
      &+
      S_{yy}\,{{\partial}_y} 
      {\rm Im}\{{A^* \,({{\partial}_y}-i q_c \varphi)\, A}\}
      +
      S_{zz} \dz {\rm Im}\{{A^* \,\dz\, A}\} 
    \end{split}
    \\
    \begin{split}
      \label{vy}  
      \rho_m\,{{\partial}_t}\,{v_y}=
      &
      \left( 
        {{\eta}_1}\,{\partial_x^2} + 
        {{\alpha}_4}\,{\partial_y^2} + 
        {{{{\alpha}_4}\,{\partial_z^2}}/{2}}
      \right) \,
      {v_y}
      \\
      &+ 
      {{\partial}_y}\,
      \left[
        \left( 
          {{\alpha}_2} + {{\eta}_1} 
        \right) \,
        {{\partial}_x}\,{v_x} + 
        {{{\alpha}_4}}\, {{\partial}_z}\,{v_z}/2 
      \right] 
      \\
      &+ 
      {{\partial}_y}\,
      \left[
        -\druck -
        {E_{{0}}}\,
        {{\epsilon}_{\perp}}\,{{\partial}_z}\,{{\Phi}_r}
        +
        2\, \alpha_2\, \zeta^E\, \dx^2\,\Phi_0
      \right] 
      +
      {{\alpha}_2}\,{{\partial}_x}\,{{\partial}_t}\,{\varphi}
      \\
      &+
      S_{xy}\,{{\partial}_x} 
      {\rm Im}\{{A^* \,({{\partial}_y}-i q_c \varphi)\, A}\}
    \end{split}
    \\
    \begin{split}
      \label{vz}
      \rho_m\,{{\partial}_t}\,{v_z}=
      & 
      \left[ 
        \left( 
          \smash[b]{\underline{{\alpha}_2}} + {{\eta}_1} 
        \right) \,{\partial_x^2} + 
        {{\alpha}_4}\,
        \left( {\partial_y^2}/2 + 
          {\partial_z^2} 
        \right)  
      \right] \,
      {v_z}
      \\
      &+ 
      {{\partial}_z}\,
      \left[
        \left( 
          \smash[b]{\underline{{\alpha}_3}}+{{\alpha}_2} + {{\eta}_1} 
        \right) 
        \,{{\partial}_x}\,
        {v_x} +
        {{{{\alpha}_4}\,{{\partial}_y}\,{v_y}}/{2}}
      \right]
      \\
      &+ 
      {{\partial}_z}\,
      \left[
        -\druck \phantom{.} 
        -\smash[b]{\underline{\left( 
              {k_{{11}}} - {k_{{22}}} 
              +2\, \nen\, e_+\, e_-/\epsilon_a
            \right) 
            \,{{\partial}_x}\,{{\partial}_y}\,{\varphi}
            }} 
      \right]
      \\
      &- 
      \left[
        \smash[b]{\underline{
            e_+ -
            2\, \nen\, e_+
            }} -
        2\, (\smash[b]{\underline{\gamma_1}}+\alpha_2)\, \zeta^E
      \right] \,
      \dx^2\,{{\partial}_z}\,\Phi_0
      \\
      &-
      {E_{{0}}}\,
      \left[  
        ({\epsilon}_{\parallel}-\smash[b]{\underline{\epsilon_a}})
        \,{\partial_x^2} + 
        {{\epsilon}_{\perp}}\,{\partial_y^2} + 
        2\,{{\epsilon}_{\perp}}\,{\partial_z^2} 
      \right] \,
      {{{\Phi}_r}}
      {+ S_z {{\partial}_z} |A|^2 }
      \\
      &+
      S_{xz}\,{{\partial}_x} 
      {\rm Im}\{{A^* {{\partial}_z}\, A}\}
      +
      S_{zx} \dz {\rm Im}\{{A^* \,\dx\, A}\} 
    \end{split}
    \\
    \begin{split}
      \label{p}
      0=\,
      &\nabla \cdot \vec v
    \end{split}
  \end{align}
\label{mfeqs1}
\end{subequations}
\label{tab:mfgleichungen}
\end{widetext}
\pagebreak

\narrowtext

\noindent {\small (Note: \LaTeX{} is confused here, so we continue with a single column)}

\appendix

\section{Results for linear stability and coupling coefficients in 3D}
\label{sec:coefficients}

Some analytic and numeric results relating the 3D description of the
pattern dynamics to hydrodynamics are presented here, in particular
analytic approximations for all coefficients entering the the results
in Sections~\ref{sec:dd}, \ref{sec:stability}, and
\ref{sec:nusselt_numbers}.  In the second lowest Fourier approximation
the onset of dielectric EC is at
\begin{align}
  \label{ec}
  E_c^2=&{\frac{4\,\omega\,{{\sigma}_{\parallel}}\, \left(
        {{\alpha}_2^2} - {{\gamma}_1}\,{{\eta}_1} \right) \, }{X \,
      \left( {{\alpha}_2} -
        \displaystyle\frac{2\,\omega\,{{\sigma}_{\parallel}}\,\left(
            {{\alpha}_2}\, {{\epsilon}_{{\parallel}}} +
            {{\epsilon}_a}\,{{\eta}_1} \right) }{
          4\,{{\epsilon}_\parallel^2}\,{\omega^2} +
          {{\sigma}_\parallel^2}} \right) -
      {{\epsilon}_a}\,{{\eta}_1}\, \,{{\sigma}_{\perp}}}}
\end{align}
with a critical wavenumber
\begin{multline}
  \label{qc}
  q_c^2=
  {\frac{2\,\omega\, \left( -{{\alpha}_2^2} + 
        {{\gamma}_1}\,{{\eta}_1} \right)
      }{{k_{{33}}}\,{{\eta}_1}}}
  \\
   \times
   \frac{
    2\,{{\epsilon}_\parallel^2}\,{\omega^2} + 
    {{\sigma}_\parallel^2}
    -
    \displaystyle
    \frac{
      2\,{{\epsilon}_a}\,{{\epsilon}_{{\parallel}}}\,
      {{\epsilon}_{\perp}}\,
      {{\eta}_1}\,{\omega^2}\,{{\sigma}_{\parallel}}
      }{
      {{\alpha}_2}\,X - 
      {{\epsilon}_a}\,{{\eta}_1}\,{{\sigma}_{\perp}}
      }
    }{
    4\,{{\epsilon}_\parallel^2}\,{\omega^2} + 
    {{\sigma}_\parallel^2}
    -
    \displaystyle
    \frac{
      2 \,X
      \,
      \omega\,{{\sigma}_{\parallel}}\,
      \left( {{\alpha}_2}\,{{\epsilon}_{{\parallel}}} + 
        {{\epsilon}_a}\,{{\eta}_1} \right)
      }{
      {{\alpha}_2} X - 
      {{\epsilon}_a}\,{{\eta}_1}\,{{\sigma}_{\perp}}
      }
    }
  ,
\end{multline}
where $X:=\epsilon_\parallel \sigma_a - \epsilon_a \sigma_\parallel$.
The accurate numerical result for $E_c^2$ \textit{vs.}\ {}$\omega$ is
nearly a perfect straight line (see Fig.~\ref{fig:ec}), which can
probably be understood by means of the approximation used in
\cite{dvgpp}.  Formula~(\ref{ec}) nicely estimates the offset of this
line for intermediate $\omega\tau_0$, but gives a different slope as
$\omega\tau_0\to\infty$.  The deviations at the lower end are an
artifact occurring with all truncated Fourier approximations.  For
$q_c^2$ the situation is similar.

In the second lowest Fourier approximation, the excitation of the
hydrodynamic fields in the convection pattern near threshold (critical
mode) is
\begin{align}
  \label{eigenvector}
  \begin{split}
    \Phi=&|A|\,\sin (q_c\,x+\arg A)\,
    \\ &\times
    \left( {{\Phi}_u} + 2\,{{\Phi}_c}\,\cos (2\,\omega\,t) - 
     2\,{{\Phi}_s}\,\sin (2\,\omega\,t) \right) ,
  \end{split}
  \\
  \begin{split}
    n_z=&2\,|A|\,\cos (q_c\,x+\arg A)\,\left( \cos (\omega\,t) + \sin
    (\omega\,t) \right) ,
  \end{split}
  \\
  \begin{split}
    f=&  2\,|A|\,\sin (q_c\,x+\arg A)\,
    \\ &\times
    \left( {f_c}\,\cos (\omega\,t) - {f_s}\,\sin (\omega\,t) \right),
  \end{split}
\end{align}
where $f$ generates a velocity field $\vec v=(\dx \dz,\dy
\dz,-\dx^2-\dy^2) f$ and the real constants $f_c$, $f_s$, $\Phi_u$,
$\Phi_c$, $\Phi_s$ are given by
\begin{gather}
  \label{ev_abbrevs}
  \begin{split}
  4 {{\eta}_1}  {f_c} {q_c^3} = 
   -4 {{\alpha}_2} \omega + {\frac{{X} {E_c^2} 
        \left( 8 {{\epsilon}_\parallel^2} {\omega^2} - 
          2 {{\epsilon}_{{\parallel}}} \omega {{\sigma}_{\parallel}} + 
          3 {{\sigma}_\parallel^2} \right) }{4 
         {{\epsilon}_\parallel^2} {\omega^2} 
         {{\sigma}_{\parallel}} + {{\sigma}_\parallel^3}}},
  \end{split}
  \\
  \begin{align}
    \begin{split}
      4 {{\alpha}_2} {f_s} {q_c^3} =& -{{\epsilon}_a} {E_c^2} - 4
      {{\gamma}_1} \omega + 4 {k_{{33}}} {q_c^2}\\
      &+ {\frac{{{\epsilon}_a} {E_c^2} \left[ 2 {{\epsilon}_a} \omega
            \left( 2 {{\epsilon}_{{\parallel}}} \omega -
              {{\sigma}_{\parallel}} \right) + {{\sigma}_a} \left( 2
              {{\epsilon}_{{\parallel}}} \omega + {{\sigma}_{\parallel}}
            \right) \right] }{4 {{\epsilon}_\parallel^2} {\omega^2} +
          {{\sigma}_\parallel^2}}},
    \end{split}
  \end{align}
\\
  \begin{split}
    {{{{\Phi}_u}}=
      {{\frac{{E_c} {{\sigma}_a}}{{q_c} {{\sigma}_{\parallel}}}}}},
  \end{split}
  \\
  \begin{split}
    {{\Phi}_c} + i {{\Phi}_s}=& {\frac{\left( 1 - i \right)  {E_c} 
        \left( 2 i {{\epsilon}_a} \omega + {{\sigma}_a} \right)
        }{2\; q_c \;(2  i{{\epsilon}_{{\parallel}}} \omega  +
           {{\sigma}_{\parallel}})}}.
  \end{split}
\end{gather}

With the help of the real constants $\adfiNull$, $\adfiEins$,
$\adfiZwei$, $\adnz$ which characterize the adjoint eigenvector

\begin{gather}
  \begin{align}
    \label{adfi0}
  \adfiNull = {\frac{{E_c}\,\left( {{\alpha}_2}\,{{\epsilon}_{{\parallel}}} + 
         {{\epsilon}_a}\,{{\eta}_1} \right) \,{q_c}}{2\,{{\alpha}_2}\,
       {{\sigma}_{\parallel}}}},
  \end{align}
  \\
  \begin{align}
    \label{adfi12}
  &\adfiEins + i\,\adfiZwei = \frac{q_c}{4\,{{\alpha}_2}\,{X}\,{E_c}\,\omega}\,\times
      \notag\\      
       &\quad
       \bigl\{ 
      +{E_c^2}\,\left[ - {{\alpha}_2}\,X + 
        {{\epsilon}_a}\,{{\eta}_1}\, \left( 2\,i\,{{\epsilon}_{\perp}}\,\omega + 
          {{\sigma}_{\perp}} \right)  \right]  
      \\
      \notag
      &\quad\quad 
       -4\,\left[ \left(\alpha_2^2  - \gamma_1\, \eta_1\right)\,\omega + 
         {k_{{33}}}\,{{\eta}_1}\,{q_c^2} \right] \,
       \left( 2\,i\, {{\epsilon}_{{\parallel}}}\,\omega + 
         {{\sigma}_{\parallel}} \right)  
      \bigr\}
      ,
  \end{align}
  \\
  \begin{align}
    \label{adnz}
    \adnz = -{\frac{{{\eta}_1}\,{q_c^2}}{{{\alpha}_2}}}
  \end{align}
\end{gather}
and the normalization factor
\begin{align}
  \label{normfactor}
  4 \N=&\, 2\,({q_c^2} -\adnz)\,{{\epsilon}_a}\,E_c^2 
  \notag \\  &
  + {E_c}\,
  {q_c}\,
  \bigl[ 
    \left(
      {{\epsilon}_a}\,\adnz-\epsilon_\parallel q_c^2
    \right)
    \,\left( {{\Phi}_u} + {{\Phi}_c} + {{\Phi}_s} \right)  
    \\&\notag
    \quad- 
    2\, {{\epsilon}_a} \,\omega \,\left( \adfiEins + \adfiZwei
    \right) 
    - 
    {{\sigma}_a}\,\left( 2 
      \,\adfiNull + \adfiEins - \adfiZwei \right)\bigr] ,
\end{align}
analytic results for most coupling coefficients entering the 3D
description can be obtained:
\begin{widetext}
\begin{align}
  \begin{split}
    \label{tau}
    2\,\N\,\tau = & \, {{\epsilon}_a}\,{E_c} \,\left( 2\,\adfiNull +
      \adfiEins -  
      \adfiZwei \right)\,q_c  - 
    2 \,{{\epsilon}_{{\parallel}}}\,\left( \adfiNull\,{{\Phi}_u} +
      \adfiEins\,{{\Phi}_c} +  
      \adfiZwei\,{{\Phi}_s} \right)\,q_c^2,
  \end{split}
  \\
  \begin{split}
    \label{xixx}
     {E_c^{2}}{{\xi}_x^2} =& \\
 8{k_{{33}}}{{\eta}_1}\Bigl\{&-6{k_{{33}}}{{\eta}_1}{q_c^2}
      {{\sigma}_{\parallel}}\left( 4{\epsilon_\parallel^2}
         {{\omega}^2} + {\sigma_\parallel^2} \right)  + 
     {E_c^2}\left[ {{\epsilon}_a}{{\eta}_1}
         \left( 2{{\epsilon}_{{\parallel}}}{{\epsilon}_{\perp}}
            {{\omega}^2}{{\sigma}_{\parallel}} + 
           2{\epsilon_\parallel^2}{\omega^2}
            {{\sigma}_{\perp}} + 
           {\sigma_\parallel^2}{{\sigma}_{\perp}} \right)  - 
        {{\alpha}_2}\left( 2{\epsilon_\parallel^2}
            {\omega^2} + {\sigma_\parallel^2} \right) X \right]
      \Bigr\}\\
        \times\Bigl\{&
     8\left[ 2{{\epsilon}_{{\parallel}}}{k_{{33}}}{{\eta}_1}
         \left( {{\alpha}_2}{{\epsilon}_{{\parallel}}} - 
           {{\epsilon}_a}{{\eta}_1} \right) {\omega^2}{q_c^2} + 
        \left( {{\alpha}_2}{{\epsilon}_{{\parallel}}} + 
           {{\epsilon}_a}{{\eta}_1} \right) 
         \left( {{\alpha}_2^2} - {{\gamma}_1}{{\eta}_1} \right) 
         {\omega^2}{{\sigma}_{\parallel}} + 
        {{\alpha}_2}{k_{{33}}}{{\eta}_1}{q_c^2}
         {{\sigma}_\parallel^2} \right] X 
       + 3{{\alpha}_2^2}E_c^2{{\sigma}_{\parallel}}{X^2}
       \\
       -8{{\epsilon}_a}{k_{{33}}}{{\eta}_1^2}
      {q_c^2}&\left( 4{{\epsilon}_\parallel^2}{\omega^2} + 
        {{\sigma}_\parallel^2} \right) {{\sigma}_{\perp}} + 
     {E_c^2}\Bigl[ {{\epsilon}_a^2}{{\eta}_1^2}
         \left( 4{{\epsilon}_\perp^2}{\omega^2}
            {{\sigma}_{\parallel}} + 
           8{{\epsilon}_{{\parallel}}}{{\epsilon}_{\perp}}{\omega^2}
            {{\sigma}_{\perp}} + 
           3{{\sigma}_{\parallel}}{{\sigma}_\perp^2} \right)  - 
        2{{\alpha}_2}{{\epsilon}_a}{{\eta}_1}
         \left( 4{{\epsilon}_{{\parallel}}}{{\epsilon}_{\perp}}
            {\omega^2} + 3{{\sigma}_{\parallel}}{{\sigma}_{\perp}}
            \right) X
         \Bigr]  \Bigr\}^{-1} 
         ,
  \end{split}
  \\
  \begin{split}
    \label{xiyy}
    2\,\N\,{{\xi}_y^2} =& 
    \, 
    {{\epsilon}_{\perp}}\,{E_c}\,\left( {{\Phi}_u} + {{\Phi}_c}+
      {{\Phi}_s} \right) \,q_c  
    -
    \left( {f_c} + {f_s} \right) \, {{\alpha}_4}\,{q_c^3}   
    \\ &    
    -4\,{{\epsilon}_{\perp}} \,\omega\,\left(\adfiZwei\,{{\Phi}_c} -
      \adfiEins\,{{\Phi}_s} \right) 
    - 
    2 \,{{\sigma}_{\perp}}\,\left( \adfiNull\,{{\Phi}_u} +
      \adfiEins\,{{\Phi}_c} +  
      \adfiZwei\,{{\Phi}_s} \right)
    + {\mathrm{flexo}},
  \end{split}
  \\
  \begin{split}
    \label{xizz}
    2\,\N\,{{\xi}_z^2} =& \,
    {{\epsilon}_{\perp}}\,{E_c}\,\left( {{\Phi}_u} +
      {{\Phi}_c} + {{\Phi}_s} \right) \,q_c
    +4\, {{\alpha}_3}\,\omega 
    -    2\,\left( {f_c} + {f_s} \right) \,
      \left({{\alpha}_1} + {{\alpha}_3} + {{\alpha}_4} + {{\alpha}_5}
    + ({{\alpha}_3}/{{\alpha}_2}) \,{{\eta}_1} \right) \,
      {q_c^3} 
    \\ &    
    - 4\,{{\epsilon}_{\perp}} \,\omega \,\left(
        \adfiZwei\,{{\Phi}_c} -
        \adfiEins\,{{\Phi}_s} \right)
      -    2 \,{{\sigma}_{\perp}}\,\left( \adfiNull\,{{\Phi}_u} + \adfiEins\,{{\Phi}_c} + 
        \adfiZwei\,{{\Phi}_s} \right)
      + {\mathrm{flexo}},
  \end{split}
  \\
  \begin{split}
    \label{landaug}
    (8/3)\,\N\, g =&\, (2\, q_c^2-4\,\adnz)\,{{\epsilon}_a}\,{E_c^2} 
    +
    4\, {q_c^2}\, \left[   
       \omega\,\left( {{\alpha}_3} + {{\gamma}_2}  \right)  
       -2\,{{\epsilon}_a}\,
       \adnz\,{{\Phi}_u}\,{{\Phi}_c} 
      \right]
    \\&
    + 
      \left( {f_c} + {f_s} \right) \left[
      \left( 6\,{{\alpha}_2} + 4\,{{\alpha}_3} \right) \adnz
      - 
      4\,
      \left( {{\alpha}_1} + {{\gamma}_2} \right)\, {q_c^2}
    \right]\, {q_c^3}
    - 2\,{E_c}\,{q_c}\, {{\sigma}_a}\, \left( 3\,\adfiNull + \adfiEins - 
      2\,\adfiZwei \right)   
    \\&
    + 
    {{\epsilon}_a}{E_c} {q_c} 
    \left[ \left(5\adnz-2 q_c^2\right)
      \left( {{\Phi}_u} + 2{{\Phi}_c} \right)  - 
      4 \omega\left( 2\adfiEins + \adfiZwei \right) \right]
    +
    8 \omega q_c^2 {{\epsilon}_a} 
    \left( 
      \adfiEins{{\Phi}_u}+2 \adfiZwei{{\Phi}_c} 
      -2 \adfiEins{{\Phi}_s}\right)
    \\& 
    + 
    4\, q_c^2\, {\sigma}_a \,\left[ 
      2\,\adfiNull\,\left( {{\Phi}_u} - {{\Phi}_s} \right)  + 
      2\,\adfiEins\,{{\Phi}_c} + 
      \adfiZwei \,\left(2\, {{\Phi}_s}-{{\Phi}_u}\right) \right]
    + {\mathrm{flexo}},
  \end{split}
  \\  
  \begin{split}
    \label{Iri}
    4\, I_r+4\, i \, I_i=& \,
    \left( i \, {{\epsilon}_a}\,\omega + {{\sigma}_a} \right) \, 
    \left[ \left( 4 -2\,i \right) \,{E_c} 
      - 
      \left( 2 - 2\,i \right) \,q_c \,\left( {{\Phi}_u} + i\,{{\Phi}_c} - 
        {{\Phi}_s} \right) \right]
    \\&+ 
    {q_c^3}\,{{\epsilon}_a}\,{E_c}\,
    \left[ \left( -3 + i \right) \,{f_c} + 
      \left( 1 - i \right) \,{f_s} \right] + 2\,{{\epsilon}_{{\parallel}}}\,{q_c^4}\,\left[ {f_c}\,
      \left( {{\Phi}_u} + {{\Phi}_c} + i\,{{\Phi}_s} \right)  + 
      i\,{f_s}\,\left( {{\Phi}_u} - {{\Phi}_c} -i\, {{\Phi}_s} \right) 
    \right] ,
  \end{split}
  \\
  \begin{split}
    \label{Irix}
    4 I_{rx}+4 i I_{ix} = &\,
    {\frac{2 \,
        \left( i\,{{\epsilon}_a}\,\omega + {{\sigma}_a} \right) }
      {{{\epsilon}_a}\,{E_c}}}\times
    \\&\,\,\left[ \vphantom{\sum} 
             4 {{\gamma}_2}\left( {f_c} - {f_s} \right){q_c^2} + 
             2{{\epsilon}_a}{E_c}
          \left( {{\Phi}_u} + {{\Phi}_c} - {{\Phi}_s} \right)  - 
          {{\epsilon}_a}\left( {{\Phi}_u^2} + 
            2 {{\Phi}_c^2} +2  {{\Phi}_s^2}  \right) 
          {q_c} + 4{q_c}\left( {k_{{33}}} - {k_{{11}}}\right)
        \right] 
    \\&-
    (2-2\, i)\,\left(i\,{{\epsilon}_a} \,\omega+{{\sigma}_a} \right)  
    \left( {{\Phi}_u} + i\,{{\Phi}_c} - {{\Phi}_s} \right)
    -
     {q_c^2}\, {{\epsilon}_a}\,{E_c}\,
    \left[ \left(- 3 + i \right) \,{f_c} +(1 - i)\,{f_s} \right]
    \\&-2
     \,{{\epsilon}_{{\parallel}}}\,
    {q_c^3}  \,\left[ {f_c}\,\left( {{\Phi}_u} + {{\Phi}_c}   + i\,
        {{\Phi}_s} \right) + i {f_s}\,
      \left( {{\Phi}_u} - {{\Phi}_c} - i\,{{\Phi}_s} \right) 
      \right]
    +
    \text{flexo}  ,
  \end{split}
  \\
  \begin{split}
    \label{Gamma}
    \Gamma=&\,   
    4\,{k_{{22}}} - 4\,{k_{{33}}}
    - {{\epsilon}_a}\,({E_c}/q_c)\,
    \left( {{\Phi}_u} + {{\Phi}_c} - {{\Phi}_s} \right)   + 
    {{\epsilon}_a}\,
    \left( {{\Phi}_u^2} + 
      2\, {{\Phi}_c^2} +  2\,{{\Phi}_s^2}  \right) -
    2\,{{\alpha}_3} \,q_c\,\left( {f_c} - {f_s} \right)
    +
    \text{flexo}.
  \end{split}
  \\
  \begin{split}
    2\,{S_x} =&\, q_c\,\left[ 2\,q_c^2\,
      \left( {{\alpha}_1} + {{\alpha}_5} + {{\gamma}_2} \right)
      \,\left( {f_c} - {f_s} \right) - {{\epsilon}_a}\,{E_c}\,
      \left( {{\Phi}_u} + {{\Phi}_c} - {{\Phi}_s} \right)  + 
      q_c \,
      {{\epsilon}_{{\parallel}}}\,\left( {{\Phi}_u^2} + 
        2\, {{\Phi}_c^2} + 2\, {{\Phi}_s^2} \right)
       \right],
  \end{split}
  \\
  \begin{split}
    \label{Sz}
    2\,{S_z} =&\, 2 \,{{\alpha}_5}\,{q_c^3}\,\left( {f_c} - {f_s} \right) + 
    {{\epsilon}_a}\,{E_c}\,\left[ 2\,{E_c} -  q_c\,
      \left( {{\Phi}_u} + {{\Phi}_c} - {{\Phi}_s} \right) \right] ,
  \end{split}
  \\
  \begin{split}
    \label{Syy}
    2\,{S_{yy}} =&\, \left(  2\,{{\eta}_2}- {{\alpha}_4}\right) \,
    \left( {f_c} - {f_s} \right) \,
    {q_c^2} 
    + {q_c}\,{{\epsilon}_{\perp}}\,\left( {{\Phi}_u^2} + 
      2\, {{\Phi}_c^2} + 2\, {{\Phi}_s^2}\right) 
    +
    \text{flexo},
  \end{split}
  \\
  \begin{split}
    \label{Szz}
    2\,{S_{zz}} =&\, 2\,\left(  {{\gamma}_2}- {{\alpha}_1}\right) \,
    \left( {f_c} - {f_s} \right) \,
    {q_c^2} 
    + {q_c}\,{{\epsilon}_{\perp}}\,\left( {{\Phi}_u^2} + 
      2\, {{\Phi}_c^2} + 2\, {{\Phi}_s^2}\right) 
    +
    \text{flexo},
  \end{split}
  \\
  \begin{split}
    \label{Sxy}
    2\,{S_{xy}} =&\,
       \left( {{\alpha}_2} + {{\alpha}_5} \right) \,
       \left( {f_c} - {f_s} \right) \,
        {q_c^2}
- {{\epsilon}_a}\,{E_c}\,
       \left( {{\Phi}_u} + {{\Phi}_c} - {{\Phi}_s} \right)   + 
    {q_c}\,{{\epsilon}_{{\parallel}}}\,\left( {{\Phi}_u^2} + 
       2\, {{\Phi}_c^2} + 2\, {{\Phi}_s^2}      \right) 
    +
    \text{flexo}  ,
  \end{split}
  \\
  \begin{split}
    \label{Sxz}
    2\,{S_{xz}} =&\, 
      2\,\left( {{\gamma}_2}-  {{\alpha}_1}\right) \,
      \left( {f_c} - {f_s} \right) \,
      {q_c^2}
    + {q_c}\,{{\epsilon}_{\perp}}\,\left( {{\Phi}_u^2} + 
      2\, {{\Phi}_c^2} + 2\, {{\Phi}_s^2}\right) 
      + 4\,q_c\,\left({k_{{33}}}- {k_{{11}}} \right) 
    +
    \text{flexo}  ,
  \end{split}
\end{align}
\end{widetext}

\begin{multicols}{2}

{}

Here ``$\text{flexo}$'' stands for flexoelectric corrections, which
involve matrix inversions and are hard to express in a compact form.
For similar reasons, no formulas are given for $\kappa_x$, $S_{xx}$,
$S_{zx}$, $I_{rz}$ and $I_{iz}$.  The coefficients $\kappa_z$ vanishes
for the second lowest Fourier approximation (and $\rho_m=0$) and
$\alpha_x$ and $\beta_y+q_c \xi_y^2$ have only flexoelectric
contributions.  The expression for $\xi_{x}^2$ has a different
structure than the other formulas because it was not calculated with
the ``center manifold'' formalism but by differentiating the
determinantal condition for stability of the basic state, which is
more effective in this case.  There are indirect contributions from
excitations of $n_z$ entering $S_{xz}$, $I_{ix}$ and $I_{rx}$ [the
bracket in Eq.~(\ref{Irix}), in Eq.~(\ref{Sxz}) they cancel
favorably].  These have been calculated only in the lowest Fourier
approximation (i.e.\ {}for the non-oscillatory part of $n_z$), which
introduces a small error.

The quality of these results can be judged by comparing the values
obtained for MBBA at $\omega \tau_0=8$ with the accurate numerical
values in Table~\ref{tab:coefficients} (only the approximation for the
indirect contribution from $n_z$ entering $S_{xz}$, $I_{ix}$ and
$I_{rx}$ is retained).  In the combinations in which the results are
expressed there, they are, for fixed $\omega \tau_0$, independent of
the electric conductivity and, with the exception of $\kappa_x,
\kappa_z$, $I_{rx}$, and some contributions involving ${{\zeta}^{{\prime E}}}$,
also largely independent of $\omega\tau_0$ for $\omega\tau_0>8$.
Two exceptions, which should both be experimentally accessible, shall
be highlighted: the dynamic flexoelectric contribution to $E_c
\alpha_x$, which increases linear in $\omega \tau_0$, and the
contribution to $g$ proportional to ${\zeta^{{\prime E}}}^2$ which is
the only one which increases $\sim(\omega \tau_0)^2$ as $\omega
\tau_0\to \infty$.  The result for $g$ in the lowest Fourier
approximation
\begin{multline}
  \label{gmow1}
  g=  9.45 + 0.00252\,{e_+^2} - 
   0.00401\,\omega\,\tau_0\,{e_+}\,{{\zeta}^{{\prime E}}} \\
   - 
   5.18\EE{-5}\,\omega\,\tau_0\,{{{{\zeta}^{{\prime E}}}}^2} + 
   0.00102\,{\omega^2\,\tau_0^2}\,{{{{\zeta}^{{\prime E}}}}^2}
\end{multline}
illustrates the latter effect, although it is correct only in its
order of magnitude.  The flexoelectric contributions in
Eq.~(\ref{gmow1}) and also in Table~\ref{tab:coefficients} are
expressed in terms of $e_+:=e_1+e_3$, $e_-:=e_1-e_3$, and
${{\zeta}^{{\prime E}}}:=2 \gamma_1 \zeta^E$, in units of $10^{-12}
{\mathrm{C}}{\mathrm{m}}^{-1}$ ($3.00\EE{-5} {\mathrm{dyn}}^{1/2}$ in
Gaussian units).  Typical values measured for $e_1$ and $e_3$
are a few times that much (see the overviews in Refs.~\cite{blichi,chl}).

As a result of the approximate $\pi/4$ phase shift of the director
oscillations, $\kappa_z$ is so small that the remaining finite viscous
effect is comparable in size to the effect of finite mass density
$\rho_m$, which has been suppressed everywhere else.

\section{``Order parameter'' \textit{vs.}\ ``center manifold'' method}
\label{sec:methods}

Here, two general methods for obtaining amplitude equations are
compared.  It is shown that they give different results in the
presence of multiple homogeneous soft modes.  In this work, the
``center manifold'' method is used to derive the reduced equations.
In order to keep the formalism simple, it will be restricted to
homogeneous ($\vec q=0$) modes and their slow modulations.  The
inclusion of patterning soft modes ($|\vec q|\ne 0$) is straight
forward.

\subsection{Formal setting}
\label{sec:setting}

Let the state vector $U(\vec r)$ describe the configuration of all
relevant degrees of freedom (e.g.~hydrodynamic fields) of the system
in the (ideally) infinitely extended, $D$-dimensional $\vec r$ space.
Assume the ``microscopic equations'' to be of the form
\begin{equation}
  \label{multilinear}
  0=F(U)=L(\dt,\nabla) U(\vec r) +\text{nonlinear terms},
\end{equation}
where the linear operator $L(\dt,\nabla)$ is polynomial in $\dt$ and
$\nabla$, acting on $U(\vec r)$ locally and translation invariant in
space and time.  There are several branches $j$ of linear modes
$V_{j}(\vec q)$ which solve the generalized eigenvalue problem
\begin{equation}
  \label{eigenstates}
  L(\sigma_j(\vec q),i \vec q)V_{j}(\vec q) =0,
\end{equation}
and, with some suitable scalar product $\left<\cdot|\cdot\right>$
(which does \emph{not} contain an integration over $\vec r$), adjoint
eigenstates
\begin{equation}
  \label{adjstates}
  W_{j}(\vec q)L(\sigma_j(\vec
  q),i \vec q)=0.
\end{equation}
For some branches $j\in K$ the growth rates
$\textrm{Re}\{\sigma_j(\vec q)\}$ vanish (or are small) at their
maxima at $\vec q=0$, for the others ($j\not\in K$) they are
negatively large in the vicinity of $\vec q=0$.

\subsection{The order parameter method}
\label{sec:opmethod}

Using the ``order parameter'' method physical states are characterized by
weighted sums over slow eigenfunctions of $L(\dt,\nabla)$,
\begin{equation}
  \label{order}
  U(\vec r)=\sum_{j\in K} \int\limits_{\Omega} u_j(\vec q)
  V_{j}(\vec q)\exp(i \vec q \cdot \vec r)  d\vec q +
  R(\{u_k\}).
\end{equation}
The weights $u_j(\vec q)$ are interpreted as the Fourier transforms of
the set of ``amplitudes'' used in the reduced description.  The range of
integration $\Omega$ is a region around $\vec q=0$, large enough to
include all significant contributions from $u_j(\vec q)$ and small
enough to exclude slow (patterning) modes at large wavenumbers.  With
this ansatz slaved contributions $R(\{u_k\})$, which are fully in the
fast eigenspace of $L$, come in only at nonlinear order.  The linear
dynamics for each amplitude is simply given by
\begin{equation}
  \label{order-dyn}
  \dt u_j(\vec q)=\sigma_j(\vec q) u_j(\vec q).
\end{equation}
An inverse Fourier transform yields the linear dynamics in physical space,
which is usually simplified by truncating the Taylor expansion of
$\sigma_j(\vec q)$ in each component of $\vec q$ for small $|\vec q|$,
such that, in physical space, derivatives of $u_j$ are obtained.
        
However, the situation is different in the case of multiple slow
branches.  Then $\sigma_j(\vec q)$ is typically non-analytic in the
components of $\vec q$ (although it is analytic in $|\vec q|$).  As a
generic example, consider the linear operator
\begin{equation}
  \label{modop}
  L(\dt,\nabla)=
  \left(
    \begin{array}{cc}
      \dx^2+\dy^2 - \dt & \dx \dy \\
      \dx \dy & \dx^2+\dy^2 - a \dt
    \end{array}
  \right)
\end{equation}
with a positive parameter $a$.  There are two neutral modes at
$(q,p):=\vec q=0$.  It is easily seen that one of the two growth rates
is of the form
\begin{equation}
  \label{modsig}
  \sigma_1(q,p)=
  -{\frac{{p^4} + {p^2}\,{q^2} + {q^4}}{{p^2} + {q^2}}} + O(a),
\end{equation}
i.e.,~non-analytic at $q,p=0$.  As a result, the corresponding
amplitude equation in physical space
\begin{multline}
  \dt u_1(\vec r)=\frac{7}{8} \nabla^2 u_1(\vec r)\\ 
  +\int K(\vec r^\prime-\vec r)\,u_1(\vec r^\prime)\,d^2 r^\prime
  + O(a) +o(u_1,u_2)
\end{multline}
is nonlocal.  In polar coordinates $K(\vec r)=-(3/\pi) r^{-4} \cos 4
\varphi$.  These conclusions do not require $a$ to be small, because
the additional terms $O(a)$ depend on $a$ and they can cancel the
non-analyticity calculated here at most at particular values of $a$.
        
This transition from local basic equations to amplitude equations with
\emph{algebraically} decaying non-localities is counter-intuitive and
misleading.  This approach has the advantage that in Fourier space the
linear dynamics~(\ref{order-dyn}) is simple.  This is useful for
calculations of patterns stability involving only a few Fourier modes.

Finally, notice that a general method to reobtain local amplitude
equations from the Fourier representation~(\ref{order-dyn}), e.g.\ by
redefining the amplitudes, should not be expected.
Equation~(\ref{order-dyn}) is general enough to include even non-local
interactions in the basic equations, which certainly cannot lead to
local amplitude equations.

\subsection{The center manifold method}
\label{sec:cmmethod}

Alternatively, in the ``center manifold'' method only the slow modes
at $\vec q=0$, $V_{j}(0)$, are used for the characterization of the
physical state
\begin{equation}
  \label{reductive}
  U(\vec r)=
  \sum_{j\in K} \int\limits_{\Omega} u_j(\vec q) 
  V_{j}(0)
  \exp(i \vec q \cdot \vec r) d\vec q +
  R(\{u_k\}).
\end{equation}
The ``slow subspace'' spanned by the sum in Eq.~(\ref{reductive}) can
be extracted by the projection operator
\begin{equation}
  \label{projektor}
  P\,\cdot\,:=\sum_{j\in K} P_j\,\cdot\,,
\end{equation}
with
\begin{multline}
  \label{projectors}
  P_j f(\vec r):=
  \int\limits_{\Omega} d\vec q
  \int \frac{d \vec{r^\prime}}{(2\pi)^D} \\
  \left<W_{j}(0) 
  \right|\left.\vphantom{W_{j}}f(\vec r^\prime)\,\right>
  V_{j}(0)
  \exp(i \vec q \cdot (\vec r-\vec{r^\prime})),
\end{multline}
where it is assumed without loss of generality that the states
$W_{j}(0)$ and $V_{j}(0)$ ($j\in K$) entering $P$ form a
bi-orthonormal system.  The ``slaved'' contributions $R(\{u_k\})$
cover the remaining subspace.

The factor $V_{j}(0)$ in Eq.~(\ref{reductive}) can be pulled out of
the integral, which is then simply the inverse Fourier transform of
$u_j(\vec q)$ into physical space $u_j(\vec r)$.  It is thus justified
to define $u_j(\vec r)$ as (the local average of) the hydrodynamic
variable $\left<W_{j}|U(\vec r)\right>$.  In particular, if
$\left<W_{j}|U(\vec r)\right>$ is conserved, so is $u_j(\vec r)$.  In
such a local representation, the vicinity of the ``center manifold''
method to a multiple scale approximation would be more obvious.
Similar simplifications would also be possible for the integrals
below, but have been suppressed in order to ease the comparison with
the ``order parameter'' method.

The function $R(\{u_k\})$ is defined by the perturbative solution of
\begin{equation}
  \label{nlreductive-dyn}
  (1-P) F(U)=0,
\end{equation}
where $U$ is given by Eq.~(\ref{reductive}) and the $u_k$ are small
and vary slowly and smoothly in space and time but are otherwise
arbitrary.

At linear order in $U$, where Eq.~(\ref{nlreductive-dyn}) reduces to
\begin{equation}
  \label{reductive-dyn}
  (1-P) L(\dt,\nabla) U=0,
\end{equation}
a form
\begin{equation}
  \label{restform}
  R(\{u_k\})=
  \sum_{j\not\in K}
  \int\limits_{\Omega} 
  r_j(\vec q) 
  V_{j}(0)
  \exp(i \vec q \cdot \vec r) d\vec q
\end{equation}
with contributions $r_j(\vec q)$ only in the vicinity of $q=0$ is
sufficient.

The amplitude equations are then given by
\begin{equation}
  \label{reductive-dyn1}
  P_i L(\dt,\nabla) U=0 \quad \text{for each $i\in K$}
\end{equation}
where $R$ is eliminated from $U$ \textit{via}
Eq.~(\ref{reductive-dyn}).

When Eqs.~(\ref{reductive-dyn},\ref{reductive-dyn1}) are satisfied
with slowly varying $u_k$ and small $R(\{u_k\})$, this implies that $U$
contains no fast eigenvectors of $L(\dt,\nabla)$.  Hence the resulting
linear dynamics for $U$ is the same as the one obtained with the
``order parameter'' method, in particular $R(\{u_k\})$ is then given
by Eq.~(\ref{restform}) with
\begin{equation}
  \label{solver}
  r_j(\vec q)=-\sum\limits_{k,l} \left<W_k(\vec q)|V_j(0)\right>^{-1} 
  \left<W_k(\vec q)|V_l(0)\right> u_l(\vec q),
\end{equation}
where $l\in K$ is running over all fast modes and $j,k\not\in K$ are
running over all slow modes (for small $|\vec q|$ the matrix
$\left<W_k(\vec q)|V_j(0)\right>$ in this expression is generally a
perturbed unit matrix and readily inverted).

To see that this method yields local dynamics for the amplitudes,
split the linear operator, restricted to the subspace selected by
$Q:=(1-P)$, like
\begin{equation}
  \label{splitl}
  Q\,L(\dt,\nabla)\,Q=
  \underset{\displaystyle :=L_0}{
    \underbrace{Q\,L(0,0)\,Q}}
  +
  l(\dt,\nabla)
\end{equation}
into a part $L_0$ which is regular, and a term which is small for slow
temporal and spatial variations of the operand and polynomial in
$\dt,\nabla$.  Calling the sum on the r.s.h.\ of Eq.~(\ref{reductive})
$S(\{u_k\})$, and suppressing the arguments (such that $U=S+R$),
equation~(\ref{reductive-dyn}) becomes
\begin{equation}
  \label{reductive-dyn2}
  Q \,L \,R = Q\,L\,Q\,R=(L_0+l)\,R=-Q\,L\,S,
\end{equation}
and is solved by expanding for small $l$, i.e.
\begin{equation}
  \label{solver2}
  R=-\sum_{n=0}^\infty 
  \left(-L_0^{-1} \,l(\dt,\nabla)\right)^n \,L_0^{-1}\,
  Q\,L(\dt,\nabla)\,S.
\end{equation}
        
When eliminating $R$ from Eq.~(\ref{reductive-dyn1}) by
Eq.~(\ref{solver2}) and truncating at some power in the derivatives
(i.e.,~for slow enough variations), linear amplitude equations with
local interactions are obtained.  The extension to the nonlinear level
is straightforward (s. Section~\ref{sec:dd_derivation}).
        
It should be noticed that with this approach all modes in the kernel
of $L(0,0)$ have to be treated as ``soft modes'', some of which,
e.g.~those resulting from gauge symmetries, may not actually have
slowly relaxing modulations associated with them (see e.g.\ the
pressure mode in Sec.~\ref{sec:smderive}).
For the simple example~(\ref{modop}) the ``center manifold'' method
leads to amplitude equations identical to the basic equations.

The reason for the difference between the two approaches is that in
the multidimensional kernel of $L(0,0)$ the choice of the basis
vectors characterizing the slow modes is not unique.  While they are
fixed (with respect to the hydrodynamic variables in the ``microscopic
equations'') for the ``center manifold'' method, they point, depending
on $\vec q$, into arbitrary directions in the slow space for the
``order parameter'' method.  For the same reason, there is no-near
identity transformation mapping one representation onto the other.

%\bibliographystyle{prsty}
%\bibliography{/home/axel/bib/bibview}

\end{multicols}

\begin{widetext}
\pagebreak
\begin{table}[h]
  \begin{center}
    \begin{tabular}{cccc}
%      \toprule
      {}                
      & MBBA\footnote{Using the parameter set ``MBBA I'' from Ref.~\cite{bzk}.}
      , $2^{\mathrm{nd}}$ lowest;               
      &$4^{\mathrm{th}}$ lowest Fourier approximation              
      & Phase 5\footnote{As tabulated in Ref.~\cite{tebk} for
      30$^\circ$C, $\epsilon_\perp=5.22 \, \epsilon_0$.},  
      $4^{\mathrm{th}}$ lowest  \\
      \colrule
      {}$E_c^2/\omega$
      &$7.52\EE{9}\,{{\mathrm{V}}^2\,{\mathrm{s}}}\,{{{\mathrm{m}}^{-2}}}$
      &$7.74\EE{9}\,{{\mathrm{V}}^2\,{\mathrm{s}}}\,{{{\mathrm{m}}^{-2}}}$
      &$4.01\EE{9}\,{{\mathrm{V}}^2\,{\mathrm{s}}}\,{{{\mathrm{m}}^{-2}}}$ \\
      {}$q_c^2/\omega$
      &$1.53\EE{9}\,{{\mathrm{s}}}\,{{{{\mathrm{m}}^{-2}}}}$
      &$1.79\EE{9}\,{{\mathrm{s}}}\,{{{{\mathrm{m}}^{-2}}}}$               
      &$1.19\EE{9}\,{{\mathrm{s}}}\,{{{{\mathrm{m}}^{-2}}}}$               \\ 
      {}$\tau/\tau_0$   &$1.01$         &$1.05$         & $0.707$       \\
      {}$q_c^2 \xi_x^2$ &$0.796$        &$1.06$         & $1.89$         \\
      {}$q_c^2 \xi_y^2$ &\multicolumn{1}{c}{$1.39 + \ldots $}
      &\multicolumn{1}{l}{$1.41 - 0.0089e_-^2 + 
        0.0138{e_-}{e_+} - 0.0032{e_+^2} -$}
      & $1.26+ \ldots $\\
      {} &\multicolumn{1}{r}{}
      &\multicolumn{1}{r}{$-  0.0012{e_-}{{\zeta}^{{\prime E}}} + 
      0.0005{e_+}{{\zeta}^{{\prime E}}} - 
      0.0004{{{\zeta}^{{\prime E}}}^2}$}&\\
      {}$q_c^2 \xi_z^2$ 
      &$5.92 + \ldots$
      &$6.00 - 0.0031{e_+^2} + 0.0017{e_+}{{\zeta}^{{\prime E}}} - 
      0.0014{{{\zeta}^{{\prime E}}}^2}$     
      &$5.59$          \\
      {}$q_c \beta_y$  &$-q_c^2 \xi_y^2 + \ldots $ 
      &$-q_c^2 \xi_y^2 + 0.0022{e_-}{e_+} +
      0.0045{e_-}{{\zeta}^{{\prime E}}}$   
      &$-q_c^2 \xi_y^2 + \ldots $               \\
      {}$E_c \alpha_x$  &$\ldots$ 
      &$0.0216\,{e_+} + 0.0439\,{{\zeta}^{{\prime E}}}$   
      &$\ldots$               \\
      {}$\kappa_x/\tau_0$& $- 0.445   $ &$-0.388$   & $-0.690$              \\
      {}$\kappa_z/\tau_0$&$ \vphantom{- 3\EE{-9} {\mathrm{m}}^3\,{\mathrm{kg}}^{-1}\,}\sim\rho_m$   &$6\EE{-7} -
      3.13\EE{-9} {\mathrm{m}}^3\,{\mathrm{kg}}^{-1}\,\rho_m$ &   \\
      {}$g$              & $9.37 + \ldots $
      &$9.39 + 0.0056{e_+^2} - 0.0452{e_+}{{\zeta}^{{\prime E}}} + 
      0.0755{{{\zeta}^{\prime E}}^2}$ 
      &$9.12 + \ldots$               \\
      {}$I_r/E_c \omega \epsilon_\perp$
      {}                & $0.265$         & $0.277 $     & $0.293$   \\
      {}$I_i/E_c \omega \epsilon_\perp$
      {}                & $0.0408$        & $0.0518$     & $0.128$ \\
      {}$q_c I_{rx}/E_c \omega \epsilon_\perp$
      {}                & $-0.0229+\ldots $        
      & $-0.0312 + 6.21\EE{-5}\,{e_+^2} - 
    3.12\EE{-5}\,{e_+}\,{{\zeta}^{{\prime E}}} - 
    3.08\EE{-5}\,{{{{\zeta}^{{\prime E}}}}^2}$     
      & $0.590 + \ldots$ \\
      {}$q_c I_{ix}/E_c \omega \epsilon_\perp$
      {}                & $-0.459+\ldots $        
      & $-0.451 - 1.00\EE{-4}\,{e_+^2} + 
    5.04\EE{-5}\,{e_+}\,{{\zeta}^{{\prime E}}} + 
    4.98\EE{-5}\,{{{{\zeta}^{{\prime E}}}}^2}$     
      & $-0.502 + \ldots $ \\
      {}$q_c I_{rz}/E_c \omega \epsilon_\perp$
      {}                & $0.7115 $        
      & $0.642 $     
      & $0.992 $ \\
      {}$q_c I_{iz}/E_c \omega \epsilon_\perp$
      {}                & $0.2254 $        
      & $0.250 $     
      & $0.671 $ \\
      {}$S_E/E_c^2\epsilon_\perp$
      {}                & $0.0728$        & $0.0851$     & $0.162$     \\
      &\multicolumn{3}{c}{\vphantom{$\Big[$}Below this line, all values are
      given in units
      of $10^{-12}{\mathrm{N}}$.}\\
      {}$S_E/q_c^{2}$
      & $16.6 \pN$ 
      & $17.1 \pN$
      & $25.2 \pN$              \\
      {}$S_x/q_c^{2}$
      & $-2.23\pN$  
      & $-1.96\pN$   
      & $-3.56\pN$              \\
      {}$S_z/q_c^{2}$
      & $6.14 \pN$  
      & $7.49 \pN$     
      & $18.2 \pN$              \\
      {}$S_{xx}/q_c$
      & $-7.34\pN$  
      & $-5.95\pN$
      & $-16.4\pN$              \\
      {}$S_{yy}/q_c$
      & $8.32+\ldots\pN$  
      & \multicolumn{1}{l}{$7.07 + 0.0419\,{e_-^2} -
        0.0561\,{e_-}\,{e_+} + 0.0141\,{e_+^2} +$}
      & $10.1 +\ldots\pN$              \\
      & $ $
      & \multicolumn{1}{r}{$+ 0.0017\,{e_-}\,{{\zeta}^{{\prime E}}} - 
        0.0022\,{e_+}\,{{\zeta}^{{\prime E}}} - 
        1.\EE{-4}\,{{{{\zeta}^{{\prime E}}}}^2}\pN$}
      & $ $ \\
      {}$S_{zz}/q_c$
      & $-17.6+\ldots\pN$  
      & \multicolumn{1}{l}{$-15.7 + 0.0101\,{e_-}\,{e_+} -
        0.0101\,{{{e_+}}^2} +$}
      & $-5.49 +\ldots\pN$              \\
      & $ $
      & \multicolumn{1}{r}{$+
        0.0050\,{e_-}\,{{\zeta}^{{\prime E}}} - 
        0.0052\,{e_+}\,{{\zeta}^{{\prime E}}} - 
        1.\EE{-4}\,{{{{\zeta}^{{\prime E}}}}^2}\pN$}
      & $ $ \\
      {}$S_{xy}/q_c$
      & $9.86+\ldots\pN$  
      & $8.66 - 0.0847\,{e_-}\,{{\zeta}^{{\prime E}}} + 
   0.0285\,{e_+}\,{{\zeta}^{{\prime E}}} - 
   0.00500\,{{{{\zeta}^{{\prime E}}}}^2}\pN$     
      & $10.2 +\ldots \pN$              \\
      {}$S_{xz}/q_c$
      & $-13.7+\ldots\pN$  
      & $-11.8 - 0.0201\,{{{e_+}}^2} - 0.0102\,{e_+}\,{{\zeta}^{{\prime E}}} - 
   1.\EE{-4}\,{{{{\zeta}^{{\prime E}}}}^2}\pN$     
      & $0.317+\ldots \pN$              \\
      {}$S_{zx}/q_c$
      & $34.8\pN$  
      & $34.8\pN$     
      & $88.8\pN$              \\
      {}$\Gamma$        
      & \multicolumn{1}{c}{$-14.6 + \ldots $} 
      & \multicolumn{1}{l}{$-14.5 - 0.557{e_-^2} +
        0.323{e_-}{e_+} - 0.0565{e_+^2} -$}
      & $-32.1 + \ldots$\\  
      {}                & \multicolumn{1}{r}{}  
      & \multicolumn{1}{r}{$- 0.195{e_-}{{\zeta}^{{\prime E}}} + 
   0.0665{e_+}{{\zeta}^{{\prime E}}} - 
   0.0100{{{\zeta}^{\prime E}}^2}$} &               \\
%      \botrule
    \end{tabular}
    \caption{Coupling coefficients in the 3D dynamics of dielectric
      EC patterns at
      $\omega\tau_0:=\omega\epsilon_\perp/\sigma_\perp=8$.  The first
      column reproduces the analytic results
      Eqs.~(\ref{ec}-\ref{Gamma}). The second and third column give
      accurate numerical results for MBBA and Phase 5 (Merck).  For
      the units of the flexoelectric constants, see
      Appendix~\ref{sec:coefficients}.  Ellipsis stand for suppressed
      flexoelectric contributions. }
    \label{tab:coefficients}
  \end{center}
\end{table}

\begin{figure}[h]
  \begin{center}
    \epsfig{file=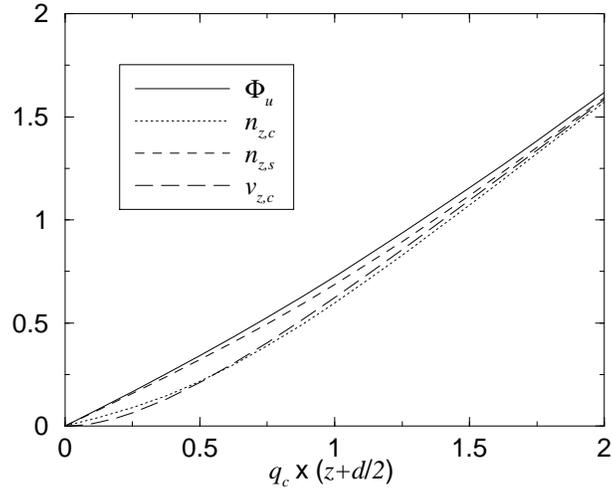,width=8 cm}
     \caption{Linear boundary layer calculated in the lowest Fourier
       approximation for MBBA.  The temporally unmodulated component
       of the potential $\Phi_u$, the in phase and out of phase
       components of the director tilt $n_{z,c}$, $n_{z,s}$, and the
       in phase component of the velocity $v_{z,c}$ are shown,
       normalization to unit slope at large distance ($\lambda \ll
       (z+d/2) \ll d$) from the boundaries.  }
    \label{fig:randschicht}
  \end{center}
\end{figure}

\begin{figure}[h]
  \begin{center}
    \epsfig{file=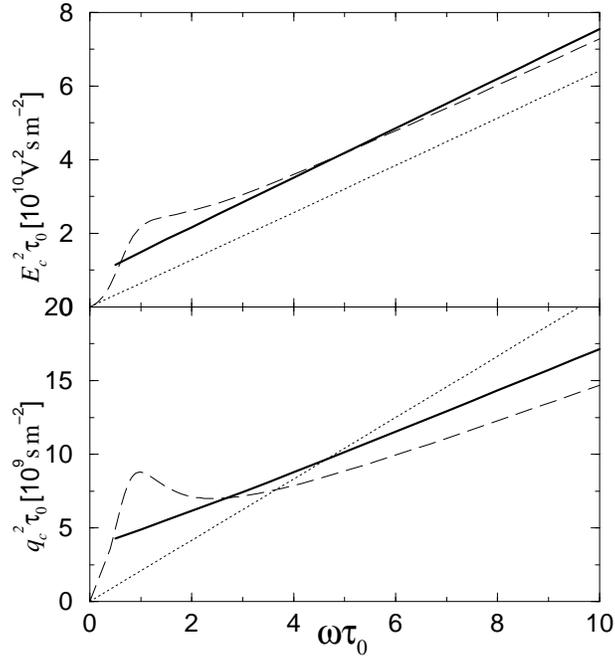,width=8 cm}
    \caption{ $E_c^2\tau_0$ and $q_c^2\tau_0$ calculated for MBBA,
      including Fourier modes up to $\omega$ (dotted), $2 \omega$
      (dashed), and $7 \omega$ (solid).  }
    \label{fig:ec}
  \end{center}
\end{figure}

\end{widetext}
\end{document}